\documentclass[aps,prc,twocolumn,showpacs,groupedaddress,amsmath,amssymb]{revtex4}

\usepackage{graphicx}
\usepackage[english]{babel}
\usepackage{array}
\usepackage{grffile} 
\usepackage{slashed} 
\usepackage{subfigure}
\usepackage{bm}
\renewcommand{\v}[1]{\bm{ #1 }}

\begin{document}
\renewcommand{\v}[1]{\bm{ #1 }}
\newcommand{\ubar}[1]{\,\overline{u}_{#1}}
\newcommand{\usp}[1]{\,u_{#1}}
\newcommand{\vbar}[1]{\,\overline{v}_{#1}}
\newcommand{\vsp}[1]{\,v_{#1}}
\newcommand{\gd}{\,\gamma \cdot}
\newcommand{\m}[1]{\mathcal{#1}} 
\newcommand{\gf}{\,\gamma_5\,}

\newcolumntype{C}{>{$}c<{$}}
\newcolumntype{R}{>{$}r<{$}}
\newcolumntype{L}{>{$}l<{$}}


\title{A Bayesian analysis of kaon photoproduction with the Regge-plus-resonance model} 



\author{Lesley De Cruz}
\author{Jan Ryckebusch}
\email[]{Jan.Ryckebusch@UGent.be}
\author{Tom Vrancx}
\author{Pieter Vancraeyveld}
\affiliation{Department of Physics and Astronomy, Ghent University, Proeftuinstraat 86, B-9000 Gent, Belgium}

\date{\today}

\begin{abstract}
  We address the issue of unbiased model selection and propose a
  methodology based on Bayesian inference to extract physical
  information from kaon photoproduction $p(\gamma,K ^{+})\Lambda$
  data.  We use the single-channel Regge-plus-resonance (RPR)
  framework for $p(\gamma,K ^{+})\Lambda$ to illustrate the proposed
  strategy.  The Bayesian evidence $\mathcal{Z}$ is a quantitative
  measure for the model's fitness given the world's data.  We present
  a numerical method for performing the multidimensional integrals in
  the expression for the Bayesian evidence.  We use the $p(\gamma,K
  ^{+})\Lambda$ data with an invariant energy $W > 2.6$~GeV in order
  to constrain the background contributions in the RPR framework with
  Bayesian inference.  Next, the resonance information is extracted
  from the analysis of differential cross sections, single and double
  polarization observables.  This background and resonance content
  constitutes the basis of a model which is coined RPR-2011.  It is
  shown that RPR-2011 yields a comprehensive account of the kaon
  photoproduction data and provides reasonable predictions for $p(e, e
  ^{\prime} K ^{+})\Lambda$ observables.
\end{abstract}
  
\pacs{14.20.Gk, 14.40.Df, 11.55.Jy}

%
%
%

\maketitle

\section{Introduction}

How to extract the nucleon resonance $(N ^{\ast})$ content of the open
strangeness photoproduction reactions $p(\gamma,K^+)\Lambda$ is a
long-standing question. Various analyses lead to disparate outcomes
concerning the set of resonances that are likely to contribute
\cite{david-1995,ireland-2004,juliadiaz-2006,mart-1999,shklyar-2005,usov-2006,shyam-2010,PhysRevC.85.034611}. The
recent availability of abundant high-statistics data has not
profoundly changed the situation so far. This indeterminacy for the
open strangeness channel is in stark contrast to the situation for
pionic channels, where the contributing resonances can be successfully
identified by means of a partial wave analysis for invariant energies
$W < 1.8$~GeV. In open strangeness channels, this technique is less
powerful as the nonresonant, or background, contributions are
larger. The importance of background contributions calls for a
framework which accounts for resonant and nonresonant processes and
which provides a means to constrain both classes of reaction
mechanisms independently.

An efficient way of pinpointing the background amplitude involves
Regge phenomenology \cite{corthals-2006,anisovich-2011}. We will
describe the $p(\gamma,K^+)\Lambda$ reaction in the so-called
Regge-plus-resonance (RPR) model, which combines ingredients of Regge
phenomenology with elements of a typical isobar approach.  The latter
belongs to the class of tree-level effective Lagrangian models. In the
RPR framework, the background amplitude is constrained by optimizing
the adjustable parameters of a Reggeized background model to data
obtained at sufficiently high energies so that the contribution of
individual resonances is projected to become marginal
\cite{corthals-2006,sibirtsev-2007}.

Even with a properly constrained background contribution, the
identification of the contributing resonances to
$p(\gamma,K^+)\Lambda$ remains a precarious task. Adding resonances
increases the amount of adjustable parameters and improves the quality
of the fit to the data. It stands to reason that one should not add
more resonances than strictly necessary, in order to obtain a good
model. One of the guiding principles for model selection is Occam's
razor \cite{rodriguez-1999}. This principle dictates that if one has
to choose between a simple and a more complex model, all else being
equal, the simpler one should be preferred. In a realistic situation,
however, all else is \emph{not} equal, and this guiding principle
should somehow be translated to a quantitative measure which balances
between model complexity on the one hand and accuracy on the other
hand. Such a measure can be derived from first principles using
Bayesian inference. This measure, called the Bayesian evidence
$\mathcal{Z}$, evaluates the overall performance of the model while
penalizing for excessive complexity.

In recent years, much effort has been directed towards a more
comprehensive description of both electromagnetic and hadronic meson
production reactions from the nucleon within coupled-channels
frameworks \cite{juliadiaz-2006,anisovich-2011,suzuki-2009}. Ideally,
one would like to apply Bayesian inference to a state-of-the-art
dynamical coupled-channels model. However, due to the multidimensional
integrals involved in the computation of the Bayesian evidence, one is
stricken by the curse of dimensionality: at worst, the computational
cost increases exponentially with the number of adjustable
parameters. Even if the number of adjustable parameters is kept in
check, the sheer number of model evaluations required for a Monte
Carlo integration calls for a realistic model of modest complexity.
In this work, we will apply Bayesian inference to the single-channel
RPR framework for kaon photoproduction. We will consider several
variants of the RPR model and use Bayesian inference to select the
most probable model given the world's data. The RPR model has been
shown to efficiently describe the $p(\gamma,K^+)\Lambda$ observables
over a broad energy range
\cite{corthals-2006,corthals-2007b,vancraeyveld-2009a}.  The RPR
framework, as it will be used in this work, has a modest number of
adjustable parameters. The background in the RPR framework consists of
Reggeized $K ^{+}(494)$ and $K ^{\ast +}(892)$ exchange in the $t$
channel.  With these assumptions, the background part has two unknown
phases and three unknown coupling constants. In addition to the background,
the RPR model incorporates $N ^{\ast}$  in the $s$ channel.
The improved version of the RPR model, as it
will be introduced in this work, uses consistent $N ^{*}$ interaction
Lagrangians and this is an enormous asset in order to reduce the
number of coupling strengths \cite{vrancx-2011}.
For each added $N ^{\ast}$ one introduces one unknown coupling
constant for $J=\frac{1} {2}$ and two unknown coupling constants for
$J \ge \frac{3} {2}$. 

The outline of the remainder of this paper is as follows. In Section
\ref{subsec:observables} the observables and kinematics of the
$N(\gamma,K)Y$ and $N(e,e^{\prime}K)Y$ reactions are introduced. In
Section ~\ref{subsec:rproutline} we summarize the underlying
assumptions of the Regge-plus-resonance formalism used to describe
these reactions. Section~\ref{sec:bayesianinference} discusses a
Bayesian approach to model selection. The computation of the Bayesian
evidence is a high-dimensional problem which requires dedicated
numerical methods and strategies.  In
Section~\ref{subsec:evidencecompu} we provide details of these
methods.  A proof of principle of the adopted numerical strategy is
described in Section~\ref{subsec:toy}.  Bayesian methodology is
applied to determine the Reggeized background amplitude in
Section~\ref{sec:bg}. In Section~\ref{sec:res}, we determine the
optimal resonant content for $p(\gamma,K^+)\Lambda$ by evaluating a
set of 11 candidate resonances. For each of them we compute the
relative resonance probability and the results are presented in
Section~\ref{subsec:resonancecontent}. The results for the various
photoproduction observables and predictions for electroproduction
observables are presented in Sections \ref{subsec:photonresults} and
\ref{subsec:electronresults}.  A conclusion is given in
Section~\ref{sec:conclusion}.

\section{Regge-plus-resonance formalism}
\label{sec:formalism}

\subsection{Observables and kinematics}
\label{subsec:observables}
\subsubsection{Photoproduction}
The unpolarized cross section for $N(\gamma,K)Y$ has the following expression
\begin{align}
 d\sigma& =\int \frac{1}{\nu_{rel} (2 \omega) (2 E_N)} \frac{d^3\v{p}_K}{\left(2\pi\right)^3} \frac{1}{2E_K} \frac{d^3\v{p}_Y}{\left(2\pi\right)^3} \frac{1}{2E_Y} \left( 2 \pi \right)^4  \nonumber\\
 \times&\delta^{\left( 4 \right)}  \left(p_N + k - p_K - p_Y  \right) \frac{1}{4} \sum_{{\lambda_{\gamma}},\lambda_N,\lambda_Y} \left|\mathcal{M}^{\lambda_N \lambda_Y} _{\lambda_{\gamma}}\right|^2, \label{eqn:photoproduction_cs}
\end{align} 
where $ \nu_{rel} $ is the relative photon-nucleon energy, the
$p_N(E_{N},\vec{p}_{N})$, $k (\omega, \vec{k})$, $p_K (E_{K}, \vec{p}
_{K})$, and $p_Y (E_{Y}, \vec{p}_{Y})$ are the four-momenta of the
nucleon, photon, kaon, and hyperon. The $\lambda_\gamma$, $\lambda_N$
and $\lambda_Y$ denote the photon, nucleon, and hyperon
polarization. 

In the center-of-momentum (c.m.) frame, the particles' four-momenta
are defined as follows:
\begin{align}
  k^{\ast}  &= (\omega^{\ast},\v{k^{\ast}}) &    p_K^{\ast} &= (E^{\ast}_K,\v{p}^{\ast}_K) \nonumber \\
  p_N^{\ast} &= (E^{\ast}_N,-\v{k}^{\ast})  &   p_Y^{\ast} &	= (E^{\ast}_Y,\v{p}^{\ast}_Y) = (E^{\ast}_Y,-\v{p}^{\ast}_K). \label{eqn:COM-momenta}
\end{align}
The $\v{z}$-axis is the propagation direction of the incident photon, and the $\v{xz}$-plane is the reaction plane.

Inserting the c.m. momenta of Eq.~(\ref{eqn:COM-momenta}) into
Eq.~(\ref{eqn:photoproduction_cs}) yields the expression for the
unpolarized differential cross section at fixed $\left(s=W^2=\left(
    k^{\ast} + p_N^{\ast} \right)^2, t= \left(p_K^{\ast} - k^{\ast}
  \right)^2 \right)$
\begin{align}
 \frac{d\sigma}{d\Omega^{\ast}_K}  = &\, \frac{1}{64\pi^2} \frac{\left|\v{p}^{\ast}_K\right|}{\omega^{\ast}} \frac{1}{\left( E_N^{\ast} + \omega^{\ast} \right)^2} \nonumber \\
\times & \frac{1}{4} \sum_{{\lambda_{\gamma}},\lambda_N,\lambda_Y} \left|\mathcal{M}^{\lambda_N \lambda_Y} _{\lambda_{\gamma}}\right|^2, \label{eqn:photoproduction_dcs}
\end{align}
where the transition amplitude can be written as the product of the
photon polarization vector $\varepsilon^{\mu}_{{\lambda_{\gamma}}}$
and the hadronic current $\m{M}^{\lambda_N \lambda_Y}
_{\lambda_{\gamma}} = \varepsilon^{\mu}_{{\lambda_{\gamma}}}
J^{\lambda_N \lambda_Y}_{\mu}$. The hadronic current adopts the form
\begin{align}
J^{\lambda_N \lambda_Y}_{\mu} &= \ubar{\lambda_Y}^{Y}(p_Y)\;T_{\mu}\;\usp{\lambda_N}^{N}(p_N) \; , 
\label{eqn:hadronic_current}
\end{align}
where $\usp{\lambda_Y}^{Y}(p_Y)$ and $\usp{\lambda_N}^{N}(p_N)$ are
the hyperon and nucleon spinors.

The target ($T$) and recoil ($P$) asymmetries are defined as
\begin{align}
T, P = \frac{d\sigma^{\lambda_X=+\frac{1}{2}} - d\sigma^{\lambda_X=-\frac{1}{2}}}
{d\sigma^{\lambda_X=+\frac{1}{2}} +d\sigma^{\lambda_X=-\frac{1}{2}}}, \label{eqn:singlepolarization}
\end{align}
where $d\sigma \equiv \frac{d\sigma}{d\Omega_K^{\ast}}$, and $\lambda_X$ is the nucleon and hyperon spin projection on the $\v{y}$-axis, respectively. The beam asymmetry $\Sigma$ follows the definition
\begin{align}
 \Sigma = \frac{d\sigma^{\perp} - d\sigma^{\parallel}}{2d\sigma}, \label{eqn:beamasymmetry}
\end{align}
where $\sigma^{\perp}$($\sigma^{\parallel}$) refers to a linear photon polarization along the $\v{y}$($\v{x}$) axis.

Double polarization observables are defined as
\begin{align}
\frac{d\sigma^{(++)} + d\sigma^{(--)} - d\sigma^{(+-)} - d\sigma^{(-+)} }
{d\sigma^{(++)} + d\sigma^{(--)} + d\sigma^{(+-)} + d\sigma^{(-+)} },
\end{align}
where $(+-)$ is a shorthand notation for
$(\lambda_A=+s_A,\lambda_B=-s_B)$, the polarizations of the particles
$A$ and $B$ that determine the asymmetry. Beam-recoil
$p(\gamma,K^+)\Lambda$ double polarization data are available for
circularly $(C_x,C_z)$ and obliquely $(O_x,O_z)$ polarized photon
beams. These are more commonly expressed in the ``primed'' reference
frame, which is rotated about the $\v{y}=\v{y}^{\prime}$-axis over an
angle $\theta_K^{\ast}$, with $\theta_K^{\ast}$ the angle between the
incoming photon and the outgoing kaon momentum in the c.m. frame.

\subsubsection{Electroproduction}
For incoming and outgoing electron four-momenta $k_{1} ( \epsilon
_{1}, \vec{k} _ {1} )$ and $k_{2} ( \epsilon _{2}, \vec{k} _ {2} )$ 
the electroproduction cross section in the one-photon exchange
approximation (OPEA) has the following form
\begin{align} 
 d\sigma = &\int \frac{1}{\nu_{{rel}} 2 \epsilon_1 2 E_N} 
\frac{d^3\v{p}_Y}{\left(2\pi\right)^3}
\frac{1}{2E_Y}\frac{d^3\v{k}_2}{\left(2\pi\right)^3}\frac{1}{2 \epsilon_2}
\frac{d^3\v{p}_K}{\left(2\pi\right)^3}\frac{1}{2E_K} \nonumber\\
 &\times
\left( 2\pi \right)^4 \delta^{(4)} \left( p_N + k_1 - p_K -p_Y - k_2 \right) \frac{1}{4} \sum_{\lambda_i} \left|\mathcal{T}_{\lambda_i}\right|^2,\label{eqn:elcrosssection_generic}
\end{align}
where the hadronic part of the reaction is evaluated in the 
$\gamma^{\ast}N$ c.m. frame and the leptonic part in the laboratory frame
$\left( p_N^{\text{lab}} \equiv (m_N,\v{0}) \right)$. This yields the
following expression for the unpolarized differential cross section
\begin{align}
 \frac{d^3\sigma}{d\epsilon_2^{\text{lab}} d\Omega_2^{\text{lab}} d\Omega^{\ast}_K} =& \frac{1}{32 (2\pi)^5} \frac{1}{m_N} \frac{\left|\v{p}^{\ast}_K\right|}{W}\frac{\epsilon_2^{\text{lab}}}{\epsilon_1^{\text{lab}}} \nonumber \\
\times& \frac{1}{4} \sum_{\lambda_1 \lambda_2 \lambda_N \lambda_Y}{\left|\mathcal{T}^{\lambda_1 \lambda_2}_{\lambda_N \lambda_Y}\right|^2}, \label{eqn:electroproduction_dcs}
\end{align}
where $\epsilon_1^{\text{lab}}$ ($\epsilon_2^{\text{lab}}$) is the
incoming (outgoing) electron energy in the lab frame and $W$ is the
invariant energy.

In the transition amplitude $\mathcal{T}^{\lambda_1
  \lambda_2}_{\lambda_N \lambda_Y}$, $\lambda_1$ and $\lambda_2$ are
the polarizations (for high-energy electrons equal to the helicities)
of the incoming and outgoing electron. This amplitude has a leptonic
and a hadronic current, connected by a photon propagator
\begin{align}
 \m{T}^{\lambda_1 \lambda_2}_{\lambda_N \lambda_Y} = e \;
l_{\mu}^{\lambda_1\lambda_2} \;\left( \frac{-g^{\mu\nu}}{k^2}  \right) \; J^{\lambda_N\lambda_Y}_{\nu}, \label{eqn:lephadcurrent}
\end{align}
where the hadronic current $J^{\lambda_N\lambda_Y}_{\nu}$ is defined
in Eq.~(\ref{eqn:hadronic_current}), 
$l_{\mu}^{\lambda_1\lambda_2}$ is the leptonic current, and $k^2 =
(k_2-k_1)^2 = - Q ^{2}$.

Therefore, $\m{T}^{\lambda_1 \lambda_2}_{\lambda_N \lambda_Y}$ can be
linked to $\m{M}_{\lambda_{\gamma}}^{\lambda_N \lambda_Y}$ and one can
write
\begin{align}
  \m{T}^{\lambda_1 \lambda_2}_{\lambda_N \lambda_Y} = \frac{e}{Q^2}
  \sum_{{\lambda_{\gamma}}=-1,0,+1}{ (-1)^{\lambda_{\gamma}}
    L_{\lambda_{\gamma}}^{\lambda_1 \lambda_2 \ast}
    \m{M}_{\lambda_{\gamma}}^{\lambda_N\lambda_Y}}, \label{eqn:electroproductionTasM}
\end{align}
where the photon propagator was rewritten using the relation
\begin{align}
  \sum_{{\lambda_{\gamma}}=0,\pm1}{ \left( -1
    \right)^{\lambda_{\gamma}}
    \varepsilon^{\ast\mu}_{{\lambda_{\gamma}}}
    \varepsilon^{\nu}_{{\lambda_{\gamma}}} }= g^{\mu\nu} + \frac{k^\mu
    k^\nu}{Q^2}, \label{eqn:photonpropagatorsub}
\end{align}
and $Q^2 = -k^2$ is the photon virtuality.

The tensor $L_{{\lambda_{\gamma}}}^{\lambda_1 \lambda_2}$ is defined
as a contraction between the photon polarization four-vector and the
leptonic current
\begin{align}
 L_{{\lambda_{\gamma}}}^{\lambda_1 \lambda_2\ast} = l^{\lambda_1 \lambda_2}_{\mu}\varepsilon^{\ast\mu}_{\lambda_{\gamma}} . \label{eqn:leptonictensor_funcl}
\end{align}

Using Eq.~(\ref{eqn:electroproductionTasM}) in the OPEA, one can
conveniently separate the quantum electrodynamics (QED) part from the
hadronic part, by defining the two tensors
 \begin{align}
  \m{L}_{{\lambda_{\gamma}}{\lambda_{\gamma}}^{\prime}} &= \sum_{\lambda_1,\lambda_2}(-1)^{{\lambda_{\gamma}}+{\lambda_{\gamma}}^{\prime}} L_{{\lambda_{\gamma}}}^{\lambda_1 \lambda_2} \left( L_{{\lambda_{\gamma}}}^{\lambda_1 \lambda_2} \right)^{\dagger} \; , \\
 \m{H}_{{\lambda_{\gamma}}{\lambda_{\gamma}}^{\prime}} &= \sum_{\lambda_N,\lambda_Y} \m{M}_{\lambda_{\gamma}}^{\lambda_N \lambda_Y} \left( \m{M}_{{\lambda_{\gamma}}^{\prime}}^{\lambda_N \lambda_Y} \right)^{\dagger}.
 \end{align}
This allows one to replace the squared transition amplitude in Eq.~(\ref{eqn:electroproduction_dcs}) by
\begin{align}
  \sum_{\lambda_1,\lambda_2,\lambda_N,\lambda_Y} \left|
    \m{T}^{\lambda_1 \lambda_2}_{\lambda_N \lambda_Y} \right|^2=
  \frac{e^2}{Q^4} \sum_{{\lambda_{\gamma}},{\lambda_{\gamma}}^{\prime}
    = 0,\pm1}
  \m{L}_{{\lambda_{\gamma}}{\lambda_{\gamma}}^{\prime}}\m{H}_{{\lambda_{\gamma}}{\lambda_{\gamma}}^{\prime}}.
\end{align}
After this replacement, the separated cross sections or structure
functions emerge. They do not depend on the kaon azimuthal angle
$\phi^{\ast}_K$ and are defined as
\begin{align}
 \dfrac{d\sigma_T}{d\Omega^{\ast}_K} &= \chi  \left( \m{H}_{1,1} + \m{H}_{-1,-1} \right)\label{eqn:sigmaT} \; , \\
 \dfrac{d\sigma_L}{d\Omega^{\ast}_K} &= 2\chi  \m{H}_{0,0}\label{eqn:sigmaL} \; ,\\ 
 \dfrac{d\sigma_{TT}}{d\Omega^{\ast}_K} &= -\chi  \left( \m{H}_{1,-1} +\m{H}_{-1,1} \right) \; , \label{eqn:sigmaTT}\\
 \dfrac{d\sigma_{LT}}{d\Omega^{\ast}_K} &= -\chi  \left( \m{H}_{0,1} + \m{H}_{1,0} - \m{H}_{-1,0} -\m{H}_{0,-1} \right) \; ,\label{eqn:sigmaLT}
\end{align}
where $\chi = \frac{1}{ \left( 16 \pi \right)^{2} W m_N } \frac{\left|\v{p}_K^{\ast}\right|}{\left(\omega^\text{lab} - \frac{Q^2}{2m_N}\right)}$.

Expressing Eq.~(\ref{eqn:electroproduction_dcs}) in terms of the separated cross sections, we obtain
 \begin{align}
  \frac{d^3\sigma}{d\epsilon_2^{\text{lab}} d\Omega_2^{\text{lab}} d\Omega^{\ast}_K} =& \, \Gamma \left(  \dfrac{d\sigma_T}{d\Omega^{\ast}_K} + \varepsilon \dfrac{d\sigma_L}{d\Omega^{\ast}_K} + \varepsilon \dfrac{d\sigma_{TT}}{d\Omega^{\ast}_K} \cos{(2\phi^{\ast}_K)}\right. \nonumber \\
  +& \left.\sqrt{\varepsilon (1 + \varepsilon) } \dfrac{d\sigma_{LT}}{d\Omega^{\ast}_K} \cos {(\phi^{\ast}_K)} \right),
  \end{align}
in which the dependence on $\phi^{\ast}_K$ has been made explicit. The virtual photon flux 
  \begin{align}
   \Gamma =& \frac{\alpha}{2\pi^2} \frac{\epsilon_2^{\text{lab}}}{\epsilon_1^{\text{lab}}}\frac{\left(\omega_\text{lab} - \frac{Q^2}{2m_N}\right)}{Q^2}\frac{1}{1-\varepsilon},
\intertext{and the virtual photon (transverse) polarization}
  \varepsilon =& \left( 1+ \frac{2 \left|\v{k}_{\text{lab}}\right|^2}{Q^2} \tan^2\frac{\theta_e}{2} \right)^{-1},
\end{align}
are defined in terms of the
electron scattering angle $\theta_e$ and  the virtual photon
three-momentum in the lab frame $\v{k}_{\text{lab}}$.

\subsection{Regge-plus-resonance formalism}
\label{subsec:rproutline}

This section deals with the dynamics of kaon production as described
by the Regge-plus-resonance (RPR) framework introduced in
Refs.~\cite{tamara-phd,corthals-2006,corthals-2007a,corthals-2007b}. The
RPR model conjoins the economic description of high-energy data by
means of Regge phenomenology with a single-channel hadrodynamical
approach in the resonance region.  We have stressed the importance of
the background diagrams in $KY$ photoproduction for a correct
determination of the resonance parameters. In a isobar model, in which
the amplitude is described as a sum of tree-level $s$, $t$ and
$u$-channel diagrams \cite{david-1995,janssen-2001}, the determination
of the background is highly model dependent
\cite{janssen-phd}. Another issue with isobar models is there
violation of the Froissart bound \cite{donnachie-2002,froissart-1961,
  froissart-2010}. Indeed, the background amplitude of isobar models
displays a power-law $s^\alpha$ dependence at large energies where the
exponent $\alpha$ depends linearly on the spin of the exchanged
particles \cite{donnachie-2002}. The RPR approach overcomes these
shortcomings by describing the nonresonant contributions to the total
amplitude by means of Regge theory \cite{corthals-2006}.

\subsubsection{Regge background}
Guidal, Laget and Vanderhaeghen showed that the exchange of a limited
number of Regge trajectories in the $t$-channel reproduces the
high-energy, forward-angle data of both photoproduction
\cite{guidal-1997a} and electroproduction
\cite{vanderhaeghen-1998,guidal-2003,guidal-phd} of pions and kaons
off the nucleon. Along those lines, the RPR background is obtained by
Reggeizing the first materializations of the lightest kaon
trajectories, $K^+(494)$ and $K^{\ast+}(892)$. The Reggeized
amplitudes are obtained by replacing the $t$-channel Feynman
propagator by a Regge one with the appropriate signature.

The odd-spin and even-spin kaon trajectories are observed to
coincide. The measured $t$-dependence of the
$d\sigma/d\Omega_K^{\ast}$ at large $s$ does not display any
pronounced structure, and this gives additional support to the strong
degeneracy of the trajectories. Therefore, the Regge propagator
reduces to
\begin{multline}
 \mathcal{P}^K_\text{Regge} (s,t) = 
\left( \dfrac{s}{s_0}\right)^{\alpha_K(t)} \dfrac{\pi\alpha_K^{\prime}}
{\sin\left( \pi\alpha_K(t) \right)} \begin{Bmatrix}
                                                                                                                                          1\\
                                                                                                                                            e^{-i\pi\alpha_K(t)}
                                                                                                                                           \end{Bmatrix}  \\
                                                                                                                \times \dfrac{1} {\Gamma\left( 1+\alpha_K(t) \right)} \; , 
\label{eqn:ReggePropagatorfull}
\end{multline}
where $\alpha_K^{\prime}$ is the slope of the trajectory and the scale
factor $s_0$ is fixed at $1 \;\mathrm{ GeV}^{2}$.

The relative sign between the odd-spin and even-spin propagators
determines the phase ($1$ or $e^{-i\pi\alpha_K(t)}$) of
$\mathcal{P}^K_\text{Regge}$ and cannot be determined from first
principles. The issue of determining this phase by comparing model
predictions with data will be addressed in Section~\ref{sec:bg}.

For vector mesons, we obtain the proper pole positions by subtracting
the spin from the trajectories in the Regge propagator. For the
$K^{\ast+}(892)$, the resulting propagator is
\begin{multline}
 \mathcal{P}^{K^{\ast}}_\text{Regge}(s,t) = \left( \dfrac{s}{s_0}\right)^{\alpha_{K^{\ast}}(t)-\alpha_0} \dfrac{\pi\alpha_{K^{\ast}}^{\prime}}{\sin\left( \pi(\alpha_{K^{\ast}}(t)-\alpha_0) \right)} \\
 \times  \begin{Bmatrix}
      1\\
      e^{-i\pi(\alpha_{K^{\ast}}(t)-\alpha_0)}                                                                                                                                           \end{Bmatrix}                                                                                                                                           \dfrac{1} {\Gamma\left( 1 + \alpha(t)_{K^{\ast}} - \alpha_0 \right)} \; , 
\label{eqn:ReggePropagatorspin1}
\end{multline}
where $ \alpha_0 = 1$.
The employed parametrization for the $K^+(494)$ and $K^{\ast+}(892)$ trajectories is given by~\cite{janssen-phd}
\begin{align}
\alpha_K (t) &=  0.70 \; \mathrm{ GeV}^{-2} \left( t - m_K^2\right),\\
\alpha_{K^{*}} (t) &=  1 + 0.85 \; \mathrm{ GeV}^{-2} \left( t - m_{K^*}^2\right) \; .
\end{align}

\subsubsection{Gauge restoration}
\label{sub:gauge}

The $K^+(494)$ exchange diagram in the $t$-channel breaks gauge
invariance. One way of restoring it is to add the electric part of the
$s$-channel Born-diagram with the same coupling constant as in the $K^+$
exchange diagram. This procedure is also applicable for a Reggeized
$t$-channel. It turns out to be essential for a proper description of
the forward-angle differential cross sections and of the beam
asymmetries in charged pion photoproduction \cite{guidal-1997a}.

For $p(\gamma,K^+)\Lambda$, the gauge-restoring $s$-channel
contribution is pivotal to account for the plateau in the differential
cross sections at very forward kaon angles or small $|t|$ \cite{guidal-1997a}. The differential
cross section for $p(\gamma,K^+)\Lambda$ is shown in
Fig.~\ref{fig:dcs__bin-wlab__func-cos_RPR2007}.

Along similar lines, gauge invariance for electroproduction can be
restored by adopting the same $Q^2$-dependence in both the
electromagnetic coupling of the $K^+$ exchange diagram and the
electric part of the $s$-channel Born term. In practice, this implies
that a monopole kaon form factor is assigned to the electric part of
the proton exchange diagram \cite{guidal-1999}.  This procedure has
been shown to result in a reasonable prediction of the
$\sigma_L/\sigma_T$ ratio \cite{guidal-2003}.


\subsubsection{Adding resonance contributions}
\label{sub:rpr}


While Regge theory provides a fair description of meson
photoproduction observables at high energies and forward angles, there
are arguments that it can also be applied in the resonance
region. Indeed, the notion of reggeon-resonance duality states that
the amplitude should be reproduced by summing over all diagrams of a
certain channel, be it the $s$, $u$ or $t$-channel
\cite{dolen-1967}.

Even though the smooth $s$ dependence of the Regge amplitude does not
allow one to describe the structures in the resonance region, the
global trends can be fairly reproduced
\cite{guidal-2003}. Furthermore, the forward peaking of the
differential cross sections supports large contributions from the
nonresonant $t$-channel background.

Inspired by these observations, Corthals \emph{et al.}
\cite{corthals-2006, corthals-2007a,corthals-2007b,tamara-phd}
developed a hybrid model for $KY$ photoproduction dubbed 
Regge-plus-resonance (RPR). We will refer to this model as
RPR-2007. The RPR-2007 model uses amplitudes which consist of
$s$-channel resonances and Reggeized $t$-channel background
terms. This approach has also been successfully applied to the
electromagnetic production of $\pi\pi$ \cite{holvoet-phd}, as well as
$\eta$ and $\eta^{\prime}$
\cite{chiang-2002}. 

The Regge background amplitude of RPR-2007 is constrained to
above-resonance ($\sqrt{s} > 3$~GeV), forward-angle
($\cos{\theta_K^{\ast}} > 0.35$) data. By extrapolating the resulting
amplitude to smaller $\sqrt{s}$, one gets a parameter-free background
for the resonance region. The $s$-channel resonances are coherently
added to the background amplitude, resulting in a hybrid amplitude for
the resonance and high-$s$ region. RPR-2007 describes the data for
forward-angle photo- and electroproduction of $K^+\Lambda$ and
$K^{+}\Sigma^0$ \cite{corthals-2006,corthals-2007a,corthals-2007b}.
With regard to the $N^{\ast}$'s, it includes the established PDG
resonances S$_{11}(1650)$, P$_{11}(1710)$, P$_{13}(1720)$, the less
established P$_{13}(1900)$, as well as the missing D$_{13}(1900)$. The
resonance parameters of the RPR-2007 model are constrained to the
$\cos{\theta_K^{\ast}} > 0.35$ data.

\begin{figure}[htbp]
 \centering
 \includegraphics[width=\columnwidth]
{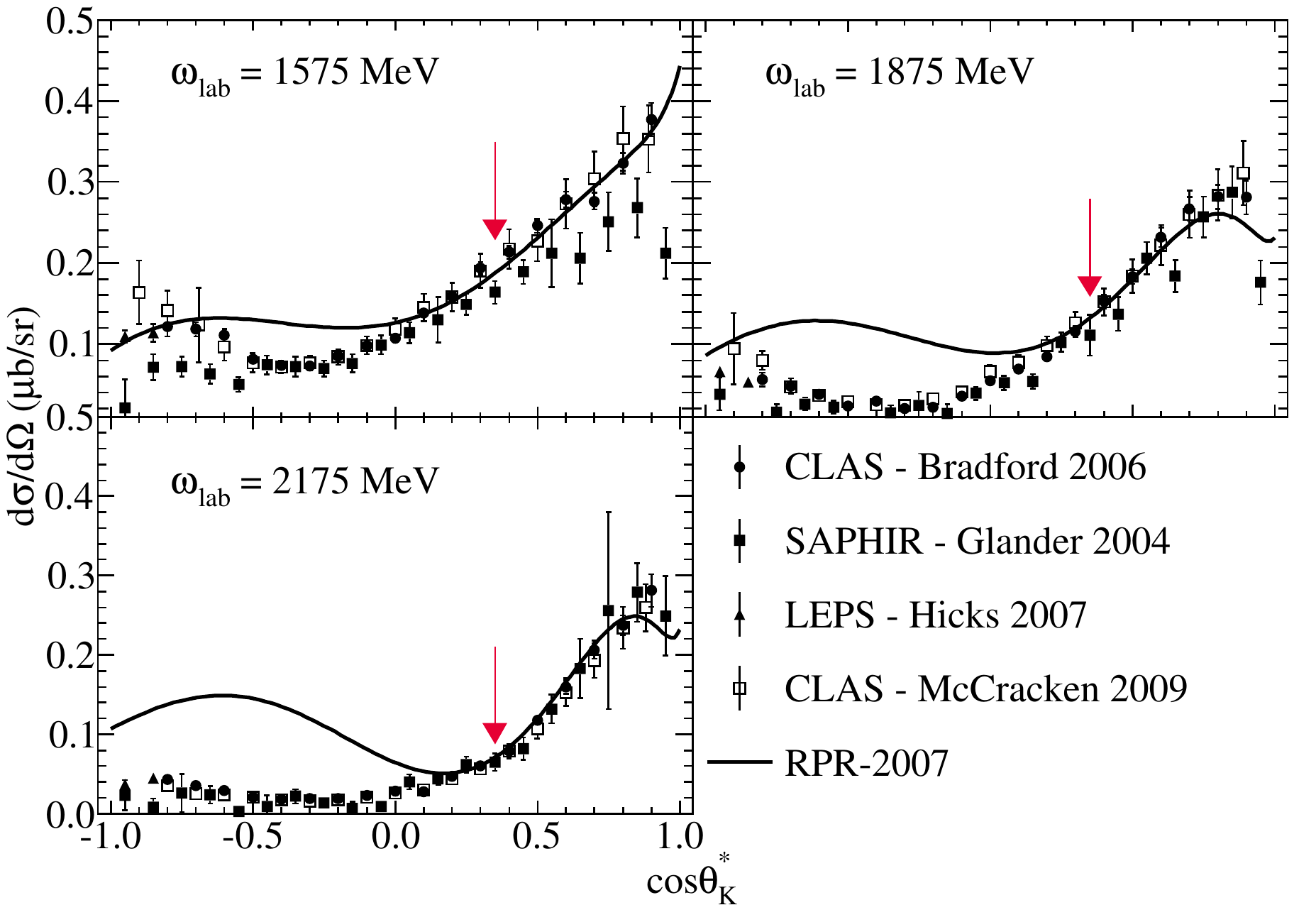}
\caption{(color online). The $ p( \gamma ,K^{+}) \Lambda$ differential
  cross section as a function of $\cos{\theta_K^{\ast}}$ for the
  laboratory photon-energy bins $\omega_{\text{lab}}=1575$ MeV,
  1875~MeV and 2175~MeV. The line denotes the RPR-2007 result and the
  data are from references
  \cite{glander-2004,hicks-2007,bradford-2006,mccracken-2009}. The
  RPR-2007 model is optimized against the $\cos{\theta_K^{\ast}} >
  0.35$ data (indicated with the arrow).}
 \label{fig:dcs__bin-wlab__func-cos_RPR2007}
\end{figure}
In Fig.~\ref{fig:dcs__bin-wlab__func-cos_RPR2007} we confront the
predictions of the RPR-2007 model with a selection of differential
cross-section data. At forward angles the data are nicely described,
in stark contrast to the situation at backward kaon angles. This
failure of the model at backward angles can be largely attributed to
the adopted description for the spin-3/2 resonance diagrams in
RPR-2007 \cite{vrancx-2011}. Obviously, unphysical bumps at backward
angles manifest themselves and the situation worsens with increasing
lab photon energy $\omega _{lab}$. In the forthcoming Section it is
pointed out how the introduction of consistent high-spin interactions can
remedy this situation.


\subsubsection{Consistent high-spin interactions}
\label{sub:consistent}

In the RPR-2007 framework, spin-3/2 resonances are described by
the Rarita-Schwinger formalism \cite{rarita-1941}. Rarita-Schwinger
fields, however, contain lower-spin components, which are not
physical. In the noninteracting Rarita-Schwinger theory these
unphysical components are eliminated by imposing the so-called
``Rarita-Schwinger constraints''. These constraints, however, do not
prevent the unphysical components from participating in the
interacting theory. The spurious lower-spin components generate
non-localities, violate causality \cite{velo-1970}, and must
therefore be avoided.

The spin-3/2 interaction Lagrangians that are used in the RPR-2007
model are inconsistent since they allow for the propagation of the
unphysical spin-1/2 modes of the Rarita-Schwinger field. These
Lagrangians involve the coupling of the spin-3/2 Rarita-Schwinger
field through the so-called ``off-shell tensor'', which contains a
free parameter. This off-shell parameter is associated with the
unphysical contribution to the spin-3/2 interaction. The spin-3/2
resonance exchange diagrams of the RPR-2007 model contain three
off-shell parameters.

In Ref.\ \cite{vrancx-2011} a consistent theory for the interaction of
high-spin fermions was devised. There it was shown that an interaction
theory that is invariant under the so-called ``unconstrained
Rarita-Schwinger gauge'' is consequently a consistent theory, i.e.\
the unphysical components of the Rarita-Schwinger field decouple from
a gauge-invariant interaction.

In the updated version of the RPR framework, dubbed RPR-2011, the
exchange of spin-3/2 resonances is described by the consistent
interaction theory of Ref.~\cite{vrancx-2011}. In addition, the RPR
model has been extended to include the exchange of spin-5/2
resonances. The expressions for the $KYR(3/2)$ and $KYR(5/2)$
interaction Lagrangians read \cite{vrancx-2011}
\begin{align}
\mathcal{L}_{KYR(3/2)} &= \frac{if_{KYR(3/2)}}{m_K^2}\overline{\Psi}^\mu_R\Gamma\psi_Y\partial_\mu\phi_K + \textrm{H.c.},\\
\mathcal{L}_{KYR(5/2)} &= -\frac{f_{KYR(5/2)}}{m_K^4}\overline{\Psi}^{\mu\nu}_R\Gamma'\psi_Y\partial_\mu\partial_\nu\phi_K + \textrm{H.c.}
\end{align}
Here, $\psi_Y$ and $\phi_K$ represent the hyperon spinor and the kaon
field. The factors $f_{KYR(3/2)}$ and $f_{KYR(5/2)}$ are strong
coupling constants. Further, $\Gamma = 1, \Gamma' = \gamma_5$ for even
parity resonances and $\Gamma = \gamma_5, \Gamma' = 1$ for odd parity
resonances. The explicitly gauge-invariant fields $\Psi^\mu_R$ and
$\Psi^{\mu\nu}_R$ describe the consistent spin-3/2 and the spin-5/2
resonances and read
\begin{align}
\Psi_R^\mu &= i(\partial^\mu \gamma_\nu \psi_R^\nu - \slashed{\partial}\psi_R^\mu),\label{Psi3/2}\\
\Psi_R^{\mu\nu} &= \partial^\mu\partial_\lambda\psi_R^{\nu\lambda} + \partial^\nu\partial_\lambda\psi_R^{\mu\lambda} - \partial^\mu\partial^\nu\gamma_\lambda\gamma_\rho\psi_R^{\lambda\rho} - \partial^2\psi_R^{\mu\nu},\label{Psi5/2}
\end{align}
where $\psi_R^\mu$ and $\psi_R^{\mu\nu}$ denote the spin-3/2 and
spin-5/2 Rarita-Schwinger fields. The interaction
Lagrangians for the $\gamma p R(3/2)$ and $\gamma p R(5/2)$ couplings
are given by
\begin{align}
\mathcal{L}_{\gamma p R(3/2)}^{(1)} &= \frac{ie\kappa^{(1)}_{pR(3/2)}}{4m_p^2}\overline{\Psi}_R^{\mu}\Gamma' \gamma^\nu \psi_p F_{\nu\mu} + \textrm{H.c.},\label{eqn:LEM32-1}\\
\mathcal{L}_{\gamma p R(3/2)}^{(2)} &= -\frac{e\kappa^{(2)}_{pR(3/2)}}{8m_p^3}\overline{\Psi}_R^{\mu}\Gamma' \partial^\nu \psi_p F_{\nu\mu} + \textrm{H.c.},\label{eqn:LEM32-2}
\end{align}
and
\begin{align}
\mathcal{L}_{\gamma p R(5/2)}^{(1)} &= -\frac{e\kappa^{(1)}_{pR(5/2)}}{16m_p^4}\overline{\Psi}_R^{\mu\nu}\Gamma \gamma^\lambda\partial_\mu \psi_p F_{\lambda\mu} + \textrm{H.c.},\label{eqn:LEM52-1} \\
\mathcal{L}_{\gamma p R(5/2)}^{(2)} &= -\frac{ie\kappa^{(2)}_{pR(5/2)}}{32m_p^5}\overline{\Psi}_R^{\mu\nu}\Gamma \partial^\lambda\partial_\mu \psi_p F_{\lambda\mu} + \textrm{H.c.}\label{eqn:LEM52-2}
\end{align}
The electromagnetic tensor $F_{\mu\nu}$ contains the photon field
$A_\mu$ and is given by $F_{\mu\nu} = \partial_\mu A_\nu
- \partial_\nu A_\mu$. Further, $\psi_p$ represents the proton spinor
and $\kappa^{(1)}_{pR(3/2)}, \kappa^{(2)}_{pR(3/2)},
\kappa^{(1)}_{pR(5/2)},$ and $\kappa^{(2)}_{pR(5/2)}$ are
electromagnetic coupling constants.

The RPR-2007 model employs a Gaussian hadronic form factor (HFF) to
regularize the transition amplitude beyond a certain energy
scale. From the expressions (\ref{Psi3/2}) and (\ref{Psi5/2}) for the
explicitly gauge-invariant fields, it is seen that the power of the
momentum dependence of a consistent interaction rises with the spin of
the exchanged particle. In Ref.\ \cite{vrancx-2011} it is shown that
unlike a Gaussian HFF a ``multidipole-Gauss form factor'' is capable
of suppressing this momentum dependence. The functional form of this
HFF reads
\begin{multline}
  F_{mG}(s;m_R,\Lambda_R,\Gamma_R,J_R) = \exp{\left( -\dfrac{(s-m_R^2)^2}{\Lambda_R^4} \right)} \\
  \times \left( \dfrac{m_R^2 \widetilde{\Gamma}_R^2(J_R)}{\left(
        s-m_R^2 \right)^2 +m_R^2\widetilde{\Gamma}_R^2(J_R)}
  \right)^{J_R-\frac{1}{2}} \; ,
\label{eqn:FmG}
\end{multline}
where $\widetilde{\Gamma}_R (J_R)$ is defined as
\begin{align}
 \widetilde{\Gamma}_R(J_R) = \dfrac{\Gamma_R}{\sqrt{2^{\frac{1}{2J_R}-1}}}.
\label{eq:valueofcutoff}
\end{align}
In this expression, $m_R, \Lambda_R, \Gamma_R,$ and $J_R$ denote the
mass, the cut-off energy, the decay width, and the spin of the
exchanged resonance. For $J_R = 1/2$, Eq.\ (\ref{eqn:FmG}) reduces to
the familiar Gaussian HFF. The RPR-2011 model uses the
multidipole-Gauss HFF of Eq.~(\ref{eqn:FmG}) in order to regularize
the high-energy behavior of the consistent spin-3/2 and spin-5/2
transition amplitudes. We use one common cut-off $\Lambda_R$ for all
resonances.

\section{Bayesian inference}
\label{sec:bayesianinference}
In this Section we outline how Bayesian inference can be used to
constrain a framework like RPR against a set of data.

\subsection{Model comparison}
Using Bayes' theorem, $P(A|B)\, P(B) = P(B|A)\,P(A)$, one can
straightforwardly derive a quantity of interest for model comparison:
the probability $P(M|\left\lbrace d_k \right\rbrace)$ of a model $M$,
given a set of experimental data $\left\lbrace d_k \right\rbrace$
\begin{equation}
P(M|\left\lbrace d_k \right\rbrace) = \frac{ P(\left\lbrace d_k \right\rbrace|M) \, P(M)}{P(\left\lbrace d_k \right\rbrace)}. \label{eqn:bayestheorem}
\end{equation}
The quantity $P(\left\lbrace d_k \right\rbrace|M)$ is referred to as
the marginal likelihood or the Bayesian evidence ($\mathcal{Z}$). If
the model $M$ can have different outcomes, which are parametrized with
a set of numbers $\bm{\alpha_M}$, marginalization yields
\begin{align}
\mathcal{Z} &\equiv P(\left\lbrace d_k \right\rbrace|M) = \int P(\left\lbrace d_k \right\rbrace,\v{\alpha_M}|M)\, d\v{\alpha_M}, \label{eqn:evidence1}\\
		&= \int \mathcal{L}(\v{\alpha_M})\, \pi(\v{\alpha_M})\, d\v{\alpha_M} \label{eqn:full-evidence}.
\end{align}
This expression states that the Bayesian evidence is the integral of
the product of two distributions: $(i)$ the probability of the dataset
$\left\lbrace d_k \right\rbrace$, given the set of parameters
$\v{\alpha_M}$ and the model $M$, and $(ii)$ the probability of the
set of parameters $\v{\alpha_M}$, given the model $M$. The first
factor, $P(\left\lbrace d_k \right\rbrace|\v{\alpha_M},M)$, can be
identified as the likelihood function,
$\mathcal{L}(\v{\alpha_M})$. Any prior knowledge of the parameters'
probability distribution before considering the data $\left\lbrace
  d_k\right\rbrace$ is contained in the second factor
$P(\v{\alpha_M}|M)$, which is referred to as the prior distribution
$\pi(\v{\alpha_M})$.

The quantity of interest for model comparison is the relative
probability of a model $M_A$ versus a model $M_B$, given the available
experimental data $\left\lbrace d_k \right\rbrace$. By applying Bayes'
theorem (\ref{eqn:bayestheorem}), the evidence ratio or Bayes factor readily
emerges from the expression for this probability ratio:
\begin{align}
 \frac{P(M_A|\left\lbrace d_k \right\rbrace)}{P(M_B|\left\lbrace d_k \right\rbrace)} &= \frac{P(\left\lbrace d_k \right\rbrace|M_A)}{P(\left\lbrace d_k \right\rbrace|M_B)} \,  \frac{P(M_A)}{P(M_B)}\label{eqn:explain_evidence}\\
	&= \frac{\mathcal{Z_A}}{\mathcal{Z_B}} \;\; \mathrm{for}\, P(M_A) = P(M_B).
\end{align}
The natural logarithm of the evidence ratio can be interpreted
qualitatively with the aid of Jeffreys' scale
\cite{jeffreys-1961,kass-1995}, given in Table \ref{tab:jeffreys}.
	\begin{table}
		\small
		\begin{center}
		\caption{Jeffreys' scale for the natural logarithms of evidence ratios $\Delta \ln\mathcal{Z} = \ln\frac{\mathcal{Z_A}}{\mathcal{Z_B}}$~\cite{jeffreys-1961, kass-1995}. It provides a translation between the evidence ratio or Bayes factor and a qualitative assessment of the premise that model $A$ is more probable than model $B$.}
		\label{tab:jeffreys}
			\begin{tabular}{rcll}
				\hline
				&$|\Delta \ln\mathcal{Z}|$&$ < 1$ 		& Not worth more than a bare mention\\
				$ 1 < $&$|\Delta \ln\mathcal{Z}|$&$ < 2.5 $	& Significant \\
				$ 2.5 < $&$|\Delta \ln\mathcal{Z}|$&$ < 5$ 	& Strong to very strong\\
				$ 5 < $&$|\Delta \ln\mathcal{Z}|$&    		& Decisive\\
				\hline
			\end{tabular}
		\end{center}
	\end{table}

\subsection{Probability of a resonance}

Bayesian inference can also be used to extract the physical properties
from the data.  For example, does the fit to a set of photoproduction
data provide evidence for the introduction of a hitherto unknown
resonance?  We present a procedure to calculate the relative
probability of a certain nucleon resonance within a model for $KY$
production, such as the RPR model. This procedure will help fill the
need for an unbiased quantity that expresses the need for introducing
an unknown resonance.

Note that all probabilities mentioned in this subsection are
implicitly conditional on a given framework.  The dependence on the
RPR framework $\mathcal{M}_{RPR}$ is implied from now on, but will be
omitted for the sake of clarity, i.e.\ $P\left( X \right) \equiv
P\left( X \; | \;\mathcal{M}_{RPR}\right)$. One can write the
probability of a given resonance $R$, given experimental data
$\left\lbrace d_k \right\rbrace$ as
\begin{align}
P\left(R \; | \left\lbrace d_k \right\rbrace \right) & = \sum_{M_i} P\left(R, M_i \;|\left\lbrace d_k \right\rbrace \right), \label{eqn:models}\\
& = \sum_{M_i} P\left(R \; | M_i, \left\lbrace d_k \right\rbrace \right) P\left(M_i \;|\left\lbrace d_k \right\rbrace \right)  \label{eqn:models2}.
\end{align}
The conditional probability $P\left(R \; | M_i, \left\lbrace d_k
  \right\rbrace \right)$ simply reduces to one if the resonance $R$ is
included in the set of resonances $S_i$ used in the model variant
$M_i$, and zero otherwise. Therefore, the summation covers only a
limited set of models,
\begin{align}
 P\left(R \; | \left\lbrace d_k \right\rbrace \right) 	& = \sum_{M_i | R \in S_i} P\left(M_i \;|\left\lbrace d_k \right\rbrace \right), \label{eqn:exclusivesum}\\
				& = \sum_{M_i | R \in S_i} P\left(\left\lbrace d_k \right\rbrace|M_i\right)  \frac{P\left(M_i\right)}{P\left(\left\lbrace d_k \right\rbrace\right)}. \label{eqn:bayesapp1}
\end{align}
Applying Bayes' theorem, one finds that the evidence
$P\left(\left\lbrace d_k \right\rbrace|M_i\right)$ appears in equation
(\ref{eqn:bayesapp1}). Assuming that there is no preference for
any specific model before comparing it to data, the factor
$\frac{P(M_i )}{P(\left\lbrace d_k \right\rbrace )}$ is equal for all
model variants $i$. Therefore, the factor can be omitted in all
subsequent calculations for the probability
ratios. This again reduces the calculation of relative probabilities $
P\left(R_1 \; | \left\lbrace d_k \right\rbrace \right)/ P\left(R_2 \;
  | \left\lbrace d_k \right\rbrace \right)$ to the evaluation of
the evidence integrals of the form of Eq.~\ref{eqn:full-evidence}.

\subsection{Likelihood function}\label{sss:likelihood}

Experimental data are usually reported to have normally distributed
errors and to be independent. The addition of $N$ squared normally
distributed, independent random variables with mean 0 and variance 1
results in a variable $X = \sum^N_{i=1} x^2_i$ that obeys a
\textit{chi-square distribution} \cite{feller-1972,spiegel-2000}
\begin{align}
f^N(X) =  \dfrac{X^{N/2-1} e^{-X/2}}{2^{N/2}\Gamma(\frac{N}{2})}. \label{eqn:chisquaredistribution}
\end{align}
The quantity $\chi^2(\v{\alpha_M})$ is defined as
\begin{equation}
\chi^2(\v{\alpha_M}) = \sum_{i=1}^N \dfrac{\left( d_i-f_i(\v{\alpha}_M)\right)^2}{\sigma_i^2},\label{eqn:chisquare}
\end{equation}
where $N$ is the total number of data points, $\sigma_i$ is the error
on data point $d_i$, and $f_i(\v{\alpha}_M)$ is the corresponding
model prediction.  The quantity $\chi^2(\v{\alpha_M})$ represents a
sum of squares of normally distributed variables and is expected to obey
the chi-square distribution of Eq.~(\ref{eqn:chisquaredistribution}).

With a likelihood function of the form  (\ref{eqn:chisquaredistribution}), we get the log-likelihood
\begin{multline}
  \ln{\mathcal{L}(\v{\alpha_M})} = \left(\frac{k}{2}-1\right) \ln{\chi^2(\v{\alpha_M})} -\frac{k}{2} \ln{2} \\
  -\ln{\Gamma \left(\frac{k}{2}\right)}-\frac{\chi^2(\v{\alpha_M})}{2} \; ,
 \label{eqn:logchisquared}
\end{multline} 
where $k$ is the number of degrees of freedom: this is equal to the
number of data points $N$ minus the number of free parameters. This
correction is necessary because by constraining the free parameters
using the data, one effectively decreases the number of degrees of
freedom.

The $\chi^2(\v{\alpha_M})$ and $\mathcal{L}(\v{\alpha_M})$ are unknown
functions of the model parameters $\v{\alpha_M}$ and the numerical
computation of the Bayesian evidence $\m{Z}$ with the aid of the
Eq.~(\ref{eqn:full-evidence}) involves a multidimensional integral
$\int d\v{\alpha}_M$ over the full parameter space. This is highly
nontrivial from the numerical point of view. In the forthcoming
section \ref{subsec:evidencecompu} we outline the adopted strategy in
order to compute the Bayesian evidence.

\subsection{Numerical computation of the Bayesian evidence}
\label{subsec:evidencecompu}
For low-dimensional problems ($d \lesssim 10$), the Nested Sampling
(NS) Monte Carlo algorithm by Skilling \cite{sivia-2006,
  skilling-2006} provides an efficient means to compute the Bayesian
evidence. The posterior distribution ${P(\v{\alpha_M}|\left\lbrace d_k
  \right\rbrace, M)}$ can also be computed by this algorithm. We
employ this method to determine the Reggeized background amplitude of
the RPR-2011 model \cite{decruz-2010}. The results of this analysis
are reported in Section~\ref{sec:bg}. 

\begin{figure}[tpb]
\begin{center}
\includegraphics[width=.9\columnwidth]{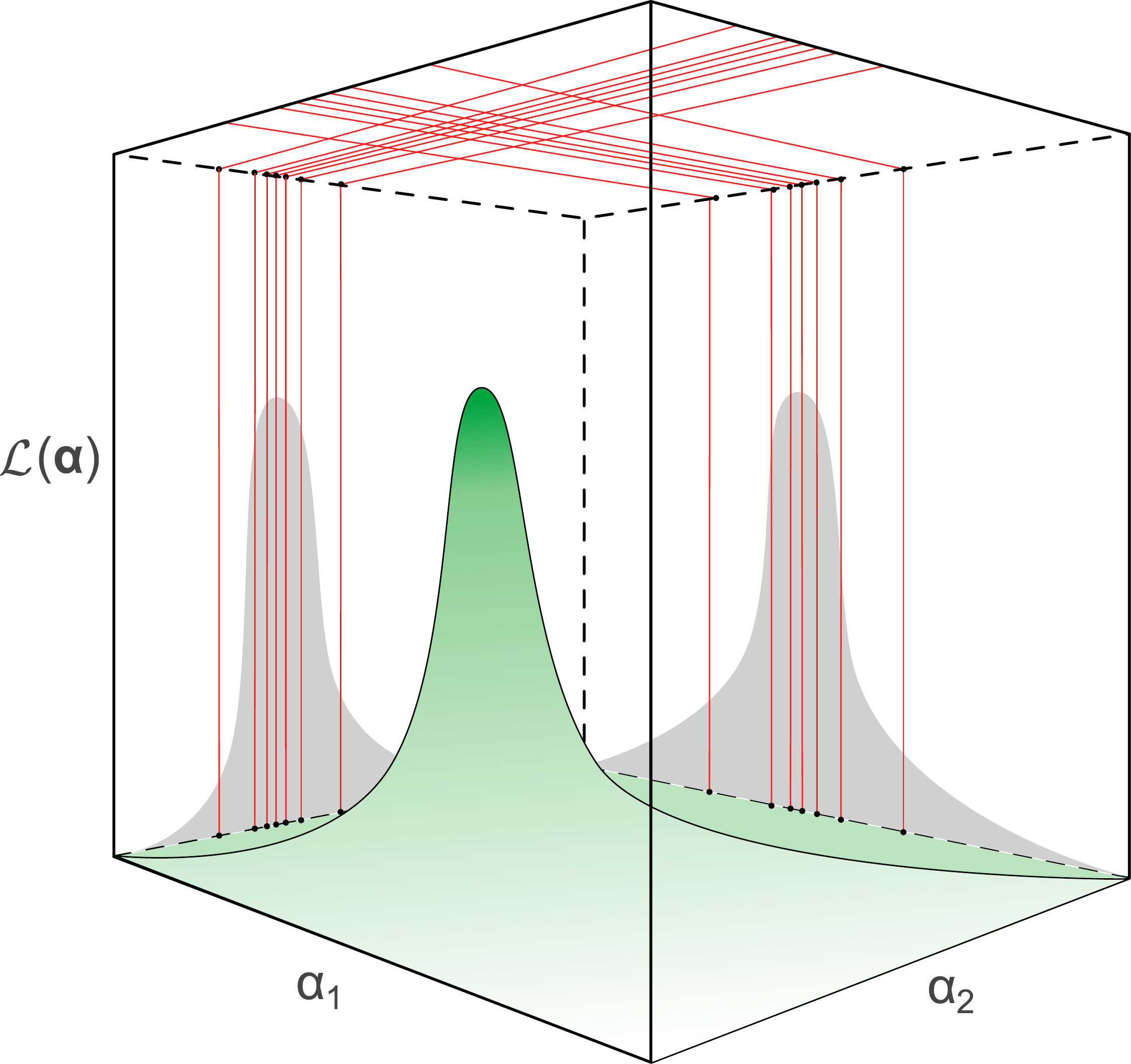}
\caption{(color online). The likelihood function $\mathcal{L}(\v{\alpha})$ (green surface) and the proposal distribution determined by {\sc vegas}, represented by a grid (red lines) with a number of bins per dimension (here 8).}
\label{fig:vegaspeak}
\end{center}
\end{figure}

In high-dimensional problems, NS has been criticized for having a
sharply decreasing acceptance rate as the likelihood constraint
becomes more exclusive \cite{chopin-2010}. Therefore, high-dimensional
problems call for an alternative numerical technique. If there is no
need to determine the posterior distribution, or if the parameters are
so-called ``nuisance parameters'', whose value are of no interest,
other Monte-Carlo integration methods can be employed. One such method
is the {\sc vegas} algorithm by Lepage \cite{lepage-1977}. {\sc vegas}
uses importance sampling: the points are sampled from a
proposal distribution which approximates the normalized integrand. The
proposal distribution is discretized in the form of an adaptive grid,
of which each cell is sampled with an equal probability. This idea is
illustrated in Fig.~\ref{fig:vegaspeak}. The {\sc vegas} algorithm is
most suitable if the integrand $\mathcal{L}(\v{\alpha _ {M}})$ can be
approximated by a separable function.

        We have adapted the GNU Scientific Library (GSL)
        implementation \cite{GSL} of the {\sc vegas} algorithm to the
        integrand of the evidence integral, which can assume very
        small values. This adapted {\sc vegas} method, which we will
        refer to as log-{\sc vegas}, requires a function which returns
        the natural logarithm of the integrand. The integral is
        computed while ensuring minimal loss of numerical accuracy
        that would occur by exponentiation of this function. This
        measure is indispensable for the integration of small
        quantities such as a likelihood.

Like any stochastic integration method, the log-{\sc vegas} algorithm
is apt to miss a highly localized maximum. We remedy this by locating
the maximum with a genetic algorithm (GA) before performing the
integration. We combine a rough search in the full parameter space
using a GA and a subsequent fine search in a selected part of the
parameter space using the gradient-based methods of {\sc minuit}, the
optimization module of the {\sc root} library
\cite{antcheva-2011}. This strategy has been successfully applied to a
precise determination of resonance parameters by Ireland \emph{et al.}
\cite{ireland-2004}.

The next step is to reduce the integration space to
the volume around the peak, with a range of the order of three
standard deviations in each dimension. The standard deviation around
the maximum can be calculated using the {\sc minos} routine of {\sc
  minuit} \cite{antcheva-2011}.

The first question that springs to mind is whether we do not risk
underestimating the evidence by limiting the integration domain to the
peak volume. We have addressed this concern by applying this method to
a toy example, which is detailed in the following Section.

\subsection{Toy example}
\label{subsec:toy}
As a proof of principle, we apply the methods outlined in
Section~\ref{subsec:evidencecompu} to a tractable and realistic-sized
problem. To this end, we use an event described by the function
$m_d(x)$ which is expressed in terms of a sum of $d$ Legendre
polynomials:
\begin{align}
 m_d(x) &= \sum_{l=0}^{d-1} a_l P_l(x) & x \in [-1,1]. \label{eqn:toymodel}
\end{align} 
The parameters $a_l (l=0,\ldots,d-1)$ are uniformly distributed in in $[-10,10]$ and
randomly generated. A mock data set with Gaussian noise is generated
from $m_d(x)$. The effectiveness of the GA is assessed by testing
whether the values $a_l$ can be determined from the
mock data. In the next step it is investigated whether Bayesian
inference can determine \textit{which model} was used to generate a
particular set of mock data. In essence, this amounts to use Bayesian
inference to find the dimension $d$ of the model from which the mock
data are generated.

We investigate the performance of a GA for models with a complexity
ranging from $d=1$ to $d=12$. We consider 4000 data points, a size
comparable to that of the world's $K^+\Lambda$ photoproduction data
set. For each data set, we attempt to determine the parameters of the
underlying model. We scale the population size in the GA linearly with
$d$. Due to its random character, convergence times can vary greatly
between the different GA runs. To account for this, we have repeated
the GA 40 times for each value of $d$, using a different, random set
of parameters for each run. We have found that convergence occurs for
all trial runs, and that the original parameters are reproduced by the
GA with an error per parameter of the order of 0.5\%.

Can one determine the model which best describes a given data set from
a number of model variants? This key question 
can be rigorously addressed using Bayesian inference. To illustrate the
potential of this method, we generate mock data using the toy model of
Eq.~(\ref{eqn:toymodel}) at a fixed $d$. In a next step, we try to
determine the underlying model (including its dimension $d$) by
calculating the Bayesian evidences for different trial models using
the log-{\sc vegas} method. This procedure is tested for data sets
generated by models of different complexity: from $d=1$ up to $d=12$.

\begin{figure*}[tbh]
 \centering
 \includegraphics[width=.7\textwidth] {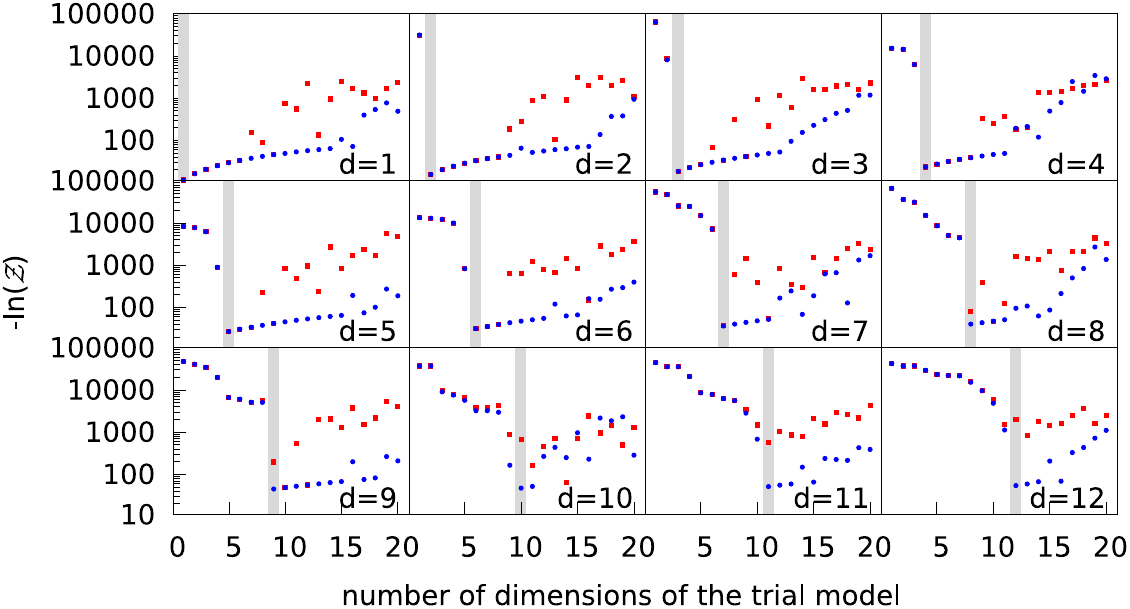}
 \caption{(color online). The $-\ln{\mathcal{Z}}$ values calculated with the log-{\sc vegas}  algorithm (red squares) and with a combined GA+{\sc minuit}+log-{\sc vegas}  integration (blue diamonds), for different model dimensions.  Each box corresponds with a model with dimension $d$, indicated with the grey band. The mock data set has 100 points. }
 \label{fig:fga_vegas_lnZ_100}
\end{figure*}

The results of the log-{\sc vegas} integrations over the entire
parameter space are compared to those limited to the peak volume in
Fig.~\ref{fig:fga_vegas_lnZ_100}. We find that the maximum evidence
value corresponds to the correct model up to at least $d = 12$.

Two striking conclusions can be drawn from
Fig.~\ref{fig:fga_vegas_lnZ_100}. A first observation is that in the
low-dimensional problems ($d \lesssim 10$), where the log-{\sc vegas}
result can be considered accurate, the two methods provide a
comparable value for the computed integrals. This means that the
likelihood in the parameter space outside the peak region is small as
compared to the maximum likelihood.  Second, the results for
high-dimensional models ($d \gtrsim 10$) indicate that the bulk of the
evidence is somehow overlooked by the global log-{\sc vegas}
integration. The global integrals for high-dimensional problems can be
orders of magnitudes smaller than those that cover only the region
around the peak. This indicates that the search space for the log-{\sc
  vegas} integration is too large in these high-dimensional problems,
and a more dedicated search strategy is required. The results of
Fig.~\ref{fig:fga_vegas_lnZ_100} indicate that a combined GA+{\sc
  minuit}+log-{\sc vegas} integration strategy is apt to the task of
dealing with high-dimensional problems.


\section{Background selection in the RPR model} 
\label{sec:bg}
In this section, we apply Bayesian inference to select the optimum
model variant for the RPR background amplitude.

\subsection{Parameters of the Reggeized background model}

 The unknown phases in
Eqs.~(\ref{eqn:ReggePropagatorfull}) and
(\ref{eqn:ReggePropagatorspin1}) give rise to several model
candidates. The possibility of the $K^{+}$ and $K^{\ast+}$
trajectories having a constant phase is excluded, as this gives rise
to a recoil asymmetry $P = 0$, which disagrees with the data. In the
forthcoming, the remaining three possibilities, namely (rotating
$K^{+}$ /rotating $K^{\ast+}$), (rotating $K^{+}$/constant $K^{\ast+}$),
and (constant $K^{+}$/rotating $K^{\ast+}$), will be referrred as RR,
RC, and CR. Apart from these variants, the background model has three
continuous parameters proportional to the product of the strong and electromagnetic couplings,
\begin{align}
e g_{K^+ \Lambda p} \qquad, & &
e\,G_{K^{\ast +}}^{v,t} = e\, g_{{K^{\ast +}}\, \Lambda p}^{v,t} \ \kappa_{K^+{K^{\ast +}}}\;.
\label{eqn:bg_free_pars}
\end{align}
Here, $ \kappa _ { K^{+} K ^{ *+}} $ is the transition magnetic moment
for $ K ^{*+} (892) \rightarrow \gamma K ^{+}(494)$ decay.  Further,
the parameters feature the strong coupling constant $g_{K^+ \Lambda
  p}$ of the $K^{+}$ trajectory and the tensor and vector couplings
$g_{{K^{\ast +}}\, \Lambda p}^{v,t}$ of the $K^{\ast +}$ trajectory.

\subsubsection{Likelihood distribution and data}
\label{subsec:likelihoodanddata}

As discussed in Section \ref{sss:likelihood}, the likelihood
distribution of the model parameters with regard to the data is the
chi-square distribution of Eq.~(\ref{eqn:chisquaredistribution}).
Recently, the CLAS collaboration published $K^+\Lambda$
\cite{mccracken-2009} and $K^+\Sigma^0$ \cite{dey-2010}
photoproduction data, featuring high-statistics differential
cross-sections and recoil polarizations. The data covers nearly the
full angular range and has 1.620~GeV $\lesssim W
\lesssim$~2.840~GeV. The broad energy range makes it a great testing
ground for both isobar, Regge and hybrid models such as RPR. Indeed,
it includes measurements taken at energies up to $W= 2.840$~GeV, which
is well above the resonance region.

\begin{figure*}[tbh]
 \centering
 \includegraphics[width=0.9\textwidth]{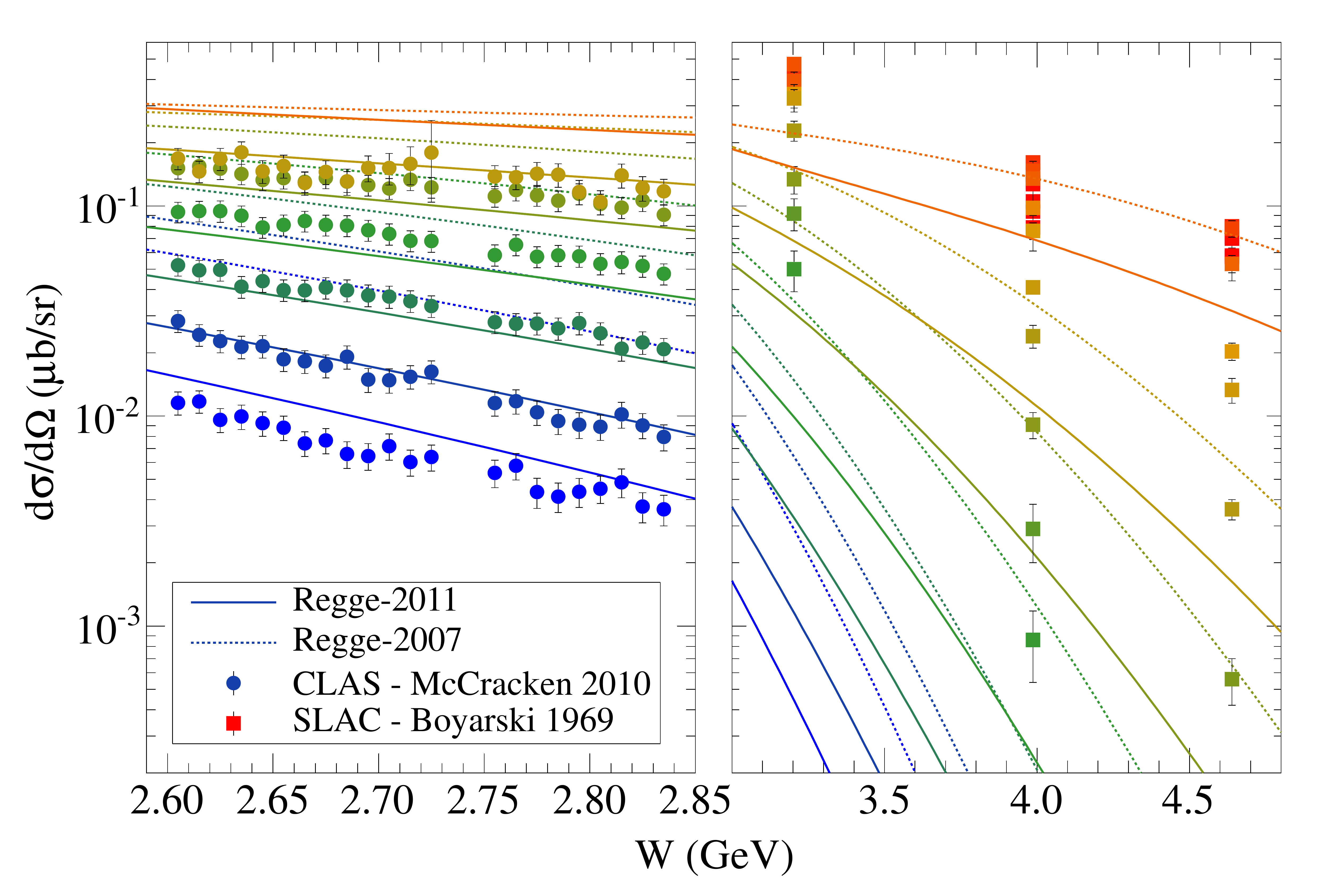} 
 \caption{(color online). The $p(\gamma,K^+)\Lambda$ differential
   cross sections as a function of $W$ for various
   $\cos\theta_K^{\ast}$. The dashed lines represent the best model
   (RR) of Ref.~\cite{decruz-2010} which follows from a Bayesian
   analysis of the $W > 3$ GeV data. The full lines correspond to the
   best model (RR) from Table~\ref{tab:lnz-iso1-clas}, optimized
   against the $2.6$ GeV $ < W < 3 $ GeV CLAS data. The lines and the
   data are color coded according to $\cos{\theta_K^{\ast}}$: from 0.4
   (blue) to 1.0 (red). The orange lines correspond to
   $\cos{\theta_K^{\ast}} = 0.95$, the other lines have a value that
   corresponds to the CLAS $\cos{\theta_K^{\ast}}$ bins, \emph{i.\ e.\
   }0.865, 0.8, 0.7, 0.6, 0.5 and 0.4.
   Data are from Refs. \cite{mccracken-2009} and
   \cite{boyarski-1969}.}
 \label{fig:mccracken}
\end{figure*}

Sibirtsev \emph{et al.} \cite{sibirtsev-2007} demonstrated that the
$p(\gamma,\pi^{+})n$ and $n(\gamma,\pi^{-})p$ reactions display
Regge-like behavior for invariant mass energies as low as 2.6~GeV.
Furthermore, Schumacher and Sargsian \cite{schumacher-2011} pointed
out that in the small-$|t|$ limit, the differential cross section for
$p(\gamma,K^{+})\Lambda$ exhibits Regge-like scaling behavior $\propto
s^{-2}$ down to $W \approx 2.3$~GeV.  One would therefore expect that
a Regge background model optimized to the $W>$~3~GeV SLAC and DESY
data \cite{decruz-2010}, provides a fair description of the
$W>$~2.6~GeV CLAS data. However, this is not the
case. Fig.~\ref{fig:mccracken} shows the $W>$~2.6~GeV $K^+\Lambda$
photoproduction data, as well as the prediction of the Reggeized
background model optimized to the $W > 3$ GeV data. Clearly, the Regge
model overshoots the CLAS data by at least a factor of 2. There is an
obvious discontinuity in the $W$ dependence between the SLAC and CLAS data
at $\cos{\theta_K^{\ast}} \approx$ 0.865, 0.8, and 0.7. 


Dey et al.~\cite{dey-2011} showed that a small set of
$p(\gamma,K^{+})\Lambda$ data from CEA \cite{elings-1967} is
inconsistent with the CLAS data. They find similar discrepancies
between new CLAS data and old high-energy data from SLAC, DESY and CEA
for other pseudoscalar meson production reactions. They conclude that
there is a persistent normalization issue in the old high-energy
differential cross-section data for a number of reactions, including
$p(\gamma,K^{+})\Lambda$ and $p(\gamma,K^{+})\Sigma^0$. The
observations of Fig.~\ref{fig:mccracken} add support to these
findings.



Because of these observations, we opt to use the data from CLAS, which
is consistent with other differential cross-section measurements in
the resonance region \cite{bradford-2006,sumihama-2006}, in order to
constrain the adjustable parameters in the Reggeized background
model. We employ the statistical methods described in
Ref.~\cite{decruz-2010}, using the 2.6~GeV $ < W < $ 3~GeV CLAS data
to compute the likelihood function. Below this energy region,
resonance contributions become more important
\cite{sibirtsev-2007}. Because the validity of Regge theory is limited
to small $|t|$, we use $\cos{\theta_K^{\ast}} > 0.35$ data to
constrain the background parameters.  With these criteria, we retain
132 differential cross sections and 130 recoil polarization $P$. This
is over a factor of four more data than for the combined SLAC/DESY
data used in the analysis reported in Ref.~\cite{decruz-2010}.


\subsubsection{Prior distribution}

We opt to use a uniform prior distribution $U$ for the coupling
constants of Eq.~(\ref{eqn:bg_free_pars}). Under conditions of highly
concentrated likelihood, compared to which the prior distribution
varies mildly, the likelihood dominates the shape of the posterior
distribution~\cite{sivia-2006}. Accordingly, the evidence calculations
will not be largely affected by the choice with regard to the shape of
the prior distribution. A sensitivity analysis will verify this
assumption.

The assumption that SU(3) symmetry is broken at the 20\% level yields
the following prior ranges for $g_{K^+\Lambda p}$
~\cite{donoghue-1982,adelseck-1990}
\begin{align}
 -4.5 \leq &\frac{g_{K^+\Lambda p}}{\sqrt{4\pi}} \leq -3.0  \; .
\label{eqn:su3predict-kl}
\end{align}
To our knowledge, for the $K^{\ast+}\Lambda p$ vertex, no reliable
theoretical constraints are available~\cite{guidal-phd}. We
choose a uniform distribution between $-$100 and +100 as the initial
prior for $\left( G_{K^{\ast +}}^{v},G_{K^{\ast +}}^{t} \right)$. To
test the sensitivity of the results to the prior width, the
calculations are repeated for a prior width of 2000 and 20000.


\subsubsection{Asymptotic behavior}
\label{sub:CLAS_asymp}

In the Regge (large $s$ and small $|t|$) limit, one can approximate
$s$ by $-u$ for fixed values of $t$. This implies that the energy
dependence of the cross section, which follows the power law
$s^{\alpha(t)}$ according to Regge theory, can be replaced by
$(\frac{s-u}{2})^{\alpha(t)}$ \cite{janssen-phd,anisovich-2010}. In an
analysis of $W>3$ GeV data, this difference is not relevant, but at
the energies considered here the difference between the two asymptotic
behaviors becomes noticeable. Therefore, we have investigated both
options using Bayesian inference.

\subsection{Results}

\subsubsection{Optimum background model variant}
\label{subsec:optimumbackground}

\begin{table*} [tbp] 
	\centering
	\caption{Logarithms of the evidence ratios ($\Delta \ln{\mathcal{Z}} \equiv \ln{\left(\mathcal{Z}/\mathcal{Z}_{max}\right)}$) for the six model variants resulting from phase combinations and asymptotic behavior options in the two-trajectory Regge model for $\gamma p \to K^{+}\Lambda$.  The prior for the coupling constant $g_{K^+\Lambda p} $ is defined by the Eq.~(\ref{eqn:su3predict-kl}). The results are listed in order of decreasing probability for the lowest prior width, $\pi = U(-100,100)$ for the $G_{K^{\ast +}}^{t,v}$ couplings. }
	\label{tab:lnz-iso1-clas}
\begin{tabular}{|c|c|c|c|c|}
\hline
	 $K^{+}$ / $K^{*+}$ phase & asymp. & { $\pi = U(-100,100)$} & 
   { $\pi = U(-1000,1000)$} & { $\pi = U(-10000,10000)$}\\
	\hline
	RR & $s$ & 0.0  &  0.0   & 0.0 \\
	RC & $s$ & $-32.7$ $\pm$ $1.4$ & 
                 $-33.5$ $\pm$ $2.7$ & 
                 $-31$  $\pm$  $13$ \\
	RR & $(s-u)/2$ & $-359.7$ $\pm$ $1.1$ & 
                 $-360.8$ $\pm$ $7.2$ & 
                 $-389$ $\pm$  $57$ \\
	RC & $(s-u)/2$ & $-432.9$ $\pm$ $1.1$ & 
                 $-435$ $\pm$ $8.9$ & 
                 $-472$ $\pm$ $58$ \\
	CR & $s$ & $-2257.2$ $\pm$ $1.1$ & 
                 $-2259.3$ $\pm$ $6.4$ & 
                 $-2282$ $\pm$  $31$ \\
	CR & $(s-u)/2$ & $-2425.5$ $\pm$ $1.1$ & 
               $-2426.3$ $\pm$ $2.6$ & 
               $-2440$ $\pm$ $27$ \\
	\hline
	\end{tabular} 
\end{table*}

\begin{table*}  
	\centering
	\caption{Logarithms of the evidence ratios ($\Delta \ln{\mathcal{Z}} \equiv \ln{\left(\mathcal{Z}/\mathcal{Z}_{max}\right)}$) for the six model variants resulting from phase combinations and asymptotic behavior options in the two-trajectory Regge model for $\gamma p \to K^{+}\Lambda$. A deviation of up to 40\% from the SU(3) prediction for $g_{K^+\Lambda p}$ is allowed. The results are listed in order of decreasing probability for the lowest prior width, $\pi = U(-100,100)$ for the $G_{K^{\ast +}}^{t,v}$ couplings.}
	\begin{tabular}{|c|c|c|c|c|}
	\hline
	 $K^{+}$ / $K^{*+}$ phase & asymp. & { $\pi = U(-100,100)$} & 
         { $\pi = U(-1000,1000)$} & { $\pi = U(-10000,10000)$}\\
	\hline
	\label{tab:lnz-iso1-clas-su3}
	RR & $s$ & 0.0  & 0.0  & 0.0 \\
	RC & $s$ & $-17.8$ $\pm$ $1.1$ & 
           $-17.4$ $\pm$ $2.9$ & 
           $-15$ $\pm$ $18$ \\
	RR & $(s-u)/2$ & $-359.7$ $\pm$ $1.0$ & 
            $-364.0$ $\pm$ $16.0$ & $-387$ $\pm$ $37$ \\
	RC & $(s-u)/2$ & $-432.7$ $\pm$ $1.2$ & $-434.9$ $\pm$ $5.3$ & 
           $-474$ $\pm$ $68$ \\
	CR & $s$ & $-2257.1$ $\pm$ $1.2$ & 
            $-2261.5$ $\pm$ $7.7$ & 
            $-2272$ $\pm$ $28$ \\
	CR & $(s-u)/2$ & $-2425.6$ $\pm$ $1.1$ & 
             $-2426.3$ $\pm$ $3.0$ & 
             $-2426$ $\pm$ $22$ \\
	\hline
	\end{tabular} 
\end{table*}

The results of our analysis are listed in
Table~\ref{tab:lnz-iso1-clas}. The data clearly favor a model
featuring an $s^{\alpha(t)}$ dependence in the cross section and two
rotating trajectories. Indeed, the difference in $\ln{\m{Z}}$ with the
second-best variant is \mbox{$32.7 \pm 1.4$}, which exceeds the value
of 5 required for a decisive statement.

The values of the coupling constants from the best model variant are
\begin{align}
	\frac{g_{K^+\Lambda p}}{\sqrt{4\pi}} & = -3.6 \pm 0.3,
	\nonumber \\ G_{K^{\ast +}}^{v} & = 9.0 \pm 0.5,
	\nonumber \\ G_{K^{\ast +}}^{t} & = 20.9 \pm 0.4.
\label{eqn:params_iso1-clasRRs}
\end{align}
In comparison with
the Bayesian analysis of Ref.~\cite{decruz-2010} which was based on the $W
>$~3~GeV data, the tensor coupling $G_{K^{\ast +}}^{t}$ has changed
sign, and its magnitude has decreased by about a factor of two. 
In Ref.~\cite{decruz-2010} which was based on the $W
> 3$~GeV data, the likelihood hypersurface exhibited a distinct
multimodal behavior, with different combinations of the coupling
constants' relative signs giving similar likelihoods.  Interestingly,
the increased amount of data used here causes the likelihood to be
concentrated in only one quadrant of the parameter space in $\left(
  G_{K^{\ast +}}^{v},G_{K^{\ast +}}^{t} \right)$ and all sign issues
for the coupling constants can be resolved. 


The expectation value (\ref{eqn:params_iso1-clasRRs}) for
$g_{K^+\Lambda p}$ is close to its SU(3) prediction of
Eq.~(\ref{eqn:su3predict-kl}). Nevertheless, we have repeated the
analyses with a prior for $g_{K^+\Lambda p}$ broader than the
condition of Eq.~(\ref{eqn:su3predict-kl}) in order to to test whether
stronger SU(3) flavor symmetry breaking is compatible with the
data. The results of this analysis are listed in Table
\ref{tab:lnz-iso1-clas-su3}.  By comparing the results of
Tables~\ref{tab:lnz-iso1-clas} and \ref{tab:lnz-iso1-clas-su3} one can
conclude that the order of the models is unaffected by the operation
of broadening the boundaries for the $g_{K^+\Lambda p}$ priors. Also
the extracted value for $g_{K^+\Lambda p}$ is not significantly
affected by the broader limits on its prior distribution. Its
expectation value becomes $ \frac{g_{K^+\Lambda p}}{\sqrt{4\pi}} =
-3.7 \pm 0.3 $ which is again compatible with the SU(3) value of
$-3.75$.

\subsubsection{High-energy predictions}

\begin{figure*}[htb]
 \centering
 \subfigure[]
  {
  \label{fig:pho-iso1-RR}
  \includegraphics[scale=0.7]{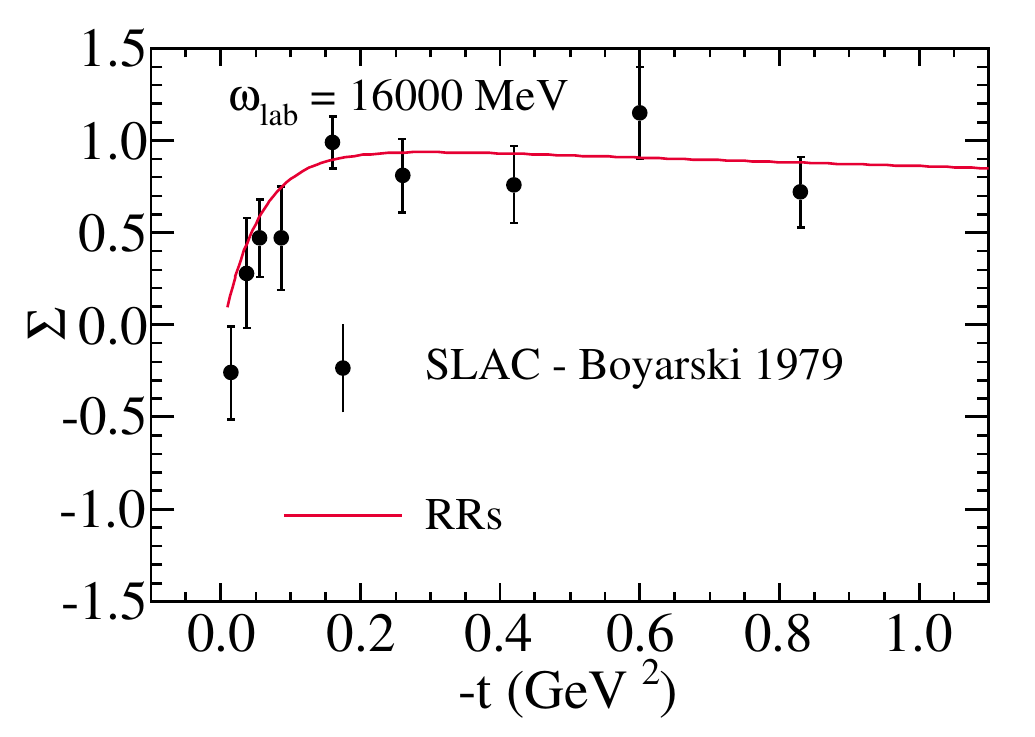}
  }
 \subfigure[ ] 
  {
  \label{fig:rec-iso1-RR}
  \includegraphics[scale=0.7]{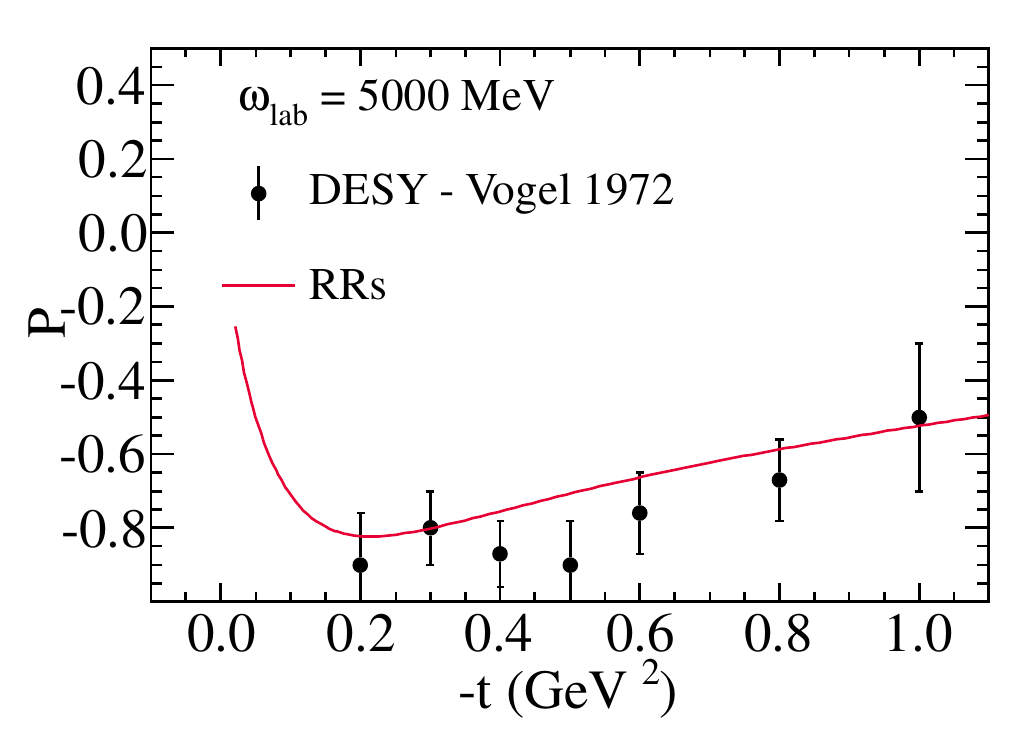}
  }
  \caption{(color online). Predictions of the best Regge model 
    from Table~\ref{tab:lnz-iso1-clas} (full red line) for the
    $p(\gamma,K^+)\Lambda$ observables $\Sigma$ and $P$ at $W > 3$
    GeV, as a function of $-t$. (a) $\Sigma$ at $\omega_{\text{lab}} =
    16$ GeV. Data are from Ref.~\cite{boyarski-1969}. (b) $P$ at
    $\omega_{\text{lab}} = 5$ GeV. Data are from
    Ref.~\cite{vogel-1972}. }
 \label{fig:pol-iso1-RR}
\end{figure*}

The high-energy differential cross section as calculated by the best
model variant for $p(\gamma,K^{+})\Lambda$ is represented by the full
lines in Fig.~\ref{fig:mccracken}. As expected, the predictions are
incompatible with the SLAC data. We attribute this to the
normalization discrepancy discussed in Ref.~\cite{dey-2011} and
Section~\ref{subsec:likelihoodanddata}. The polarization observables
$\Sigma$ \cite{boyarski-1969,quinn-1979} and $P$ \cite{vogel-1972} are
not sensitive to normalization issues. Predictions for $\Sigma$ at
$\omega_{\text{lab}} = 16$ GeV are shown in Fig.~\ref{fig:pho-iso1-RR}
for $K^{+}\Lambda$. Fig.~\ref{fig:rec-iso1-RR} shows the predictions
for $P$ at $\omega_{\text{lab}} = 5$ GeV. These predictions display an
excellent agreement with data. By constraining the Reggeized
background at $2.6$ GeV $ < W < $ 3 GeV, one can predict $P$ and
$\Sigma$ at $W > 3$ GeV. This highlights the predictive power of a
Regge model at high $W$, and corroborates the assumption that the
Reggeized background model can be constrained against $ W \gtrsim$
2.5~GeV observables.

Summarizing the background evaluations, we find that the optimum
two-trajectory Regge model for $p(\gamma,K^+)\Lambda$ features two
rotating phases and positive vector and tensor couplings. We also find
that an asymptotic $s^{\alpha(t)}$ dependence of the Regge amplitude
is preferred over a $\left( (s-u)/2 \right)^{\alpha(t)}$ one. This
model will be referred to as Regge-2011 and determines the prior for
the background amplitude of the RPR-2011 model.

\section{Resonance selection in the RPR model} 
\label{sec:res}
Given the world's $p(\gamma,K^{+})\Lambda$ data, this section
addresses the following questions: (a) From a proposed set of
resonances, what subset features in the most probable model? (b) What
is the probability of a proposed resonance $R$? Bayesian inference
allows one to answer these questions in a quantitative way.

\subsection{Data and resonances}
\label{subsec:dataandresonances}
An overview of the available $p(\gamma,K^{+})\Lambda$ data is listed
in Table \ref{tab:phodata-kl}. In view of the normalization issue
discussed in Section \ref{sec:bg}, the data from SLAC
\cite{boyarski-1969} and DESY \cite{vogel-1972} are not included in
the analysis presented below. The total number of data points which we
incorporate is 6148, of which 3455 are differential cross sections,
2241 are single and 452 are double polarization results. We stress
that after accounting for the error bars all data carry the same weight.

\begin{table}[tbh]
  \caption{Overview of the published experimental data for the reaction $p(\gamma, K^+)\Lambda$.}
\begin{center}
\begin{tabular}{|c|c|c|c|r|}
\hline
Observable & \#data & Experiment & Year & Reference\\
\hline\hline
$\frac{d\sigma}{d\Omega}$ & 56 & SLAC & 1969 & Boyarski \cite{boyarski-1969} \\ 
$$ & 720 & SAPHIR & 2004 & Glander \cite{glander-2004} \\ 
$$ & 1377 & CLAS & 2006 & Bradford \cite{bradford-2006} \\ 
$$ & 12 & LEPS & 2007 & Hicks \cite{hicks-2007} \\ 
$$ & 2066 & CLAS & 2010 & McCracken \cite{mccracken-2009} \\ 
\hline\hline
$\Sigma$ & 9 & SLAC & 1979 & Quinn \cite{quinn-1979} \\ 
$$ & 45 & LEPS & 2003 & Zegers \cite{zegers-2003} \\ 
$$ & 54 & LEPS & 2006 & Sumihama \cite{sumihama-2006} \\ 
$$ & 4 & LEPS & 2007 & Hicks \cite{hicks-2007} \\ 
$$ & 66 & GRAAL & 2007 & Lleres \cite{lleres-2007} \\ 
\hline
$T$ & 3 & BONN & 1978 & Althoff \cite{althoff-1978} \\ 
$$ & 66 & GRAAL & 2008 & Lleres \cite{lleres-2008} \\ 
\hline
$P$ & 7 & DESY & 1972 & Vogel \cite{vogel-1972} \\ 
$$ & 233 & CLAS & 2004 & McNabb \cite{mcnabb-2004} \\ 
$$ & 66 & GRAAL & 2007 & Lleres \cite{lleres-2007} \\ 
$$ & 1707 & CLAS & 2010 & McCracken \cite{mccracken-2009} \\ 
\hline\hline
$C_x \;,C_z$ & 320 & CLAS & 2007 & Bradford \cite{bradford-2007} \\ 
$O_{x^{\prime}}\; , O_{z^{\prime}}$ & 132 & GRAAL & 2008 & Lleres \cite{lleres-2008} \\ 
\hline
\end{tabular}
\label{tab:phodata-kl}
\end{center}
\end{table}

We use the differential cross section data measured by the CLAS
collaboration \cite{bradford-2006,mccracken-2009} and LEPS
\cite{hicks-2007}. Due to unresolved discrepancies with other data
sets \cite{bydzovsky-2007}, the SAPHIR differential cross-section data
\cite{glander-2004} is excluded. This decision is motivated by the
fact that the different cross-section measurements by CLAS are
internally consistent \cite{mccracken-2009} and consistent with the
LEPS data \cite{hicks-2007}. To date, there is no independent
measurement that confirms the {SAPHIR} data.

The single polarization data consists of two sets of recoil
polarization data published by the CLAS collaboration
\cite{mcnabb-2004,mccracken-2009}, as well as a set from
GRAAL~\cite{lleres-2007}. The beam asymmetry data used in our analysis
includes results from LEPS \cite{zegers-2003,sumihama-2006,hicks-2007}
and GRAAL \cite{lleres-2007}. The included target asymmetries were
determined by means of beam-recoil measurements by the GRAAL
collaboration \cite{lleres-2008}. The included double polarization
observables are beam-recoil asymmetries, consisting of $C_{x}$ and
$C_{z}$ data by CLAS \cite{bradford-2007} and GRAAL's measurements of
$O_{x^{\prime}}$ and $O_{z^{\prime}}$ \cite{lleres-2008}.

The 11 resonances considered in this work and their properties are
listed in Table~\ref{tab:resonances}. We have ``established'' as well
as ``missing'' nucleon resonances. As for their quantum numbers, mass,
width and transition form factors, we take the values quoted by the
PDG. If these are not available, we employ the values determined by
analyses based on {CQM} predictions \cite{mart-1999}. This allows us
to keep the number of adjustable parameters small.

\begin{table}[tpb]
\caption{The nucleon resonances evaluated in the analysis given in the notation $L_{2I, 2J}(M)$, along with their PDG status, spin ($J$) and parity ($\pi$), Breit-Wigner mass ($M$), width ($\Gamma$), and the uncertainty on the width ($\Delta \Gamma$).}
\label{tab:resonances}
 \centering
\begin{tabular}{|R|C|C|R|R|R|}
\hline
\text{Resonance}	& \text{ PDG status}	& J^\pi	&M \text{(MeV)}&\Gamma \text{(MeV)}& \Delta \Gamma \text{(MeV)}\\
\hline
S_{11}(1535) 	&{\star}{\star}{\star}{\star}	& 1/2^-	& 1535	& 150 & \pm 25		\\	
S_{11}(1650) 	&{\star}{\star}{\star}{\star} 	& 1/2^-	& 1650	& 150 &	\pm 20		\\	
D_{15}(1675) 	&{\star}{\star}{\star}{\star} 	& 5/2^-	& 1675	& 150 &	-20/+15		\\	
F_{15}(1680) 	&{\star}{\star}{\star}{\star} 	& 5/2^+	& 1685	& 130 &	\pm 10		\\	
D_{13}(1700) 	&{\star}{\star}{\star}		& 3/2^-	& 1700	& 100 &	\pm 50		\\	
P_{11}(1710) 	&{\star}{\star}{\star}		& 1/2^+	& 1710	& 100 &	-50/+150	\\	
P_{13}(1720) 	&{\star}{\star}{\star}{\star} 	& 3/2^+	& 1720	& 150 &	-50/+100	\\	
D_{13}(1900) 	&\text{missing}			& 3/2^-	& 1895	& 200 &	-	 	\\	
P_{13}(1900) 	&{\star}{\star}			& 3/2^+	& 1900	& 500 &	-360/+80 	\\	
P_{11}(1900) 	&\text{missing}			& 1/2^+	& 1895	& 200 &	-	 	\\	
F_{15}(2000) 	&{\star}{\star}			& 5/2^+	& 2000	& 140 &	-40/+30		\\	
\hline
\end{tabular}
\end{table}

The established four-star resonances listed by the PDG are
$S_{11}(1650)$, $D_{15}(1675)$, $F_{15}(1680)$ and $P_{13}(1720)$. The
four-star $S_{11}(1535)$ lies below the kaon production threshold, but
is included because of its large decay width and its strong predicted
coupling to the open strangeness sector \cite{an-2011}.  To our
knowledge, the contribution of the three-star $D_{13}(1700)$ to
$p(\gamma,K^{+})\Lambda$ is confirmed only by the Giessen analysis 
\cite{shklyar-2005}. The $P_{11}(1710)$, which is found in some
$K^{+}\Lambda$ analyses, is evaluated as well. The importance of this
resonance in the $\pi N$ system was questioned in the most recent
SAID analyses \cite{arndt-2006, workman-2011, said-online}. The
$P_{11}(1710)$ has also been identified in the $\pi \pi
N$ system \cite{khemchandani-2008}.

Furthermore, the two-star resonances $P_{13}(1900)$ and
$F_{15}(2000)$\footnote{The latest Review of Particle Physics
  \cite{nakamura-2010} lists this resonance with a lower mass than the
  2008 Review \cite{amsler-2008}; the new estimate is $1850-1950$
  MeV.} are evaluated. The first of these, $P_{13}(1900)$, was found
to couple to $K^{+}\Lambda$ by the Giessen group \cite{shklyar-2007}
and by the RPR-2007 model, and accounts for the structure in the
energy dependence of the differential cross-section data at $W
\approx$~1900~MeV. Schumacher and Sargsian \cite{schumacher-2011} show
that the differential cross section data from CLAS
\cite{mccracken-2009} supports one or more resonances at $W \approx 2$
GeV. Therefore, the consideration of the $F_{15}(2000)$ seems
justified. The missing $D_{13}(1900)$ and $P_{11}(1900)$ resonances
earlier introduced in the Ghent isobar model
\cite{janssen-2001,janssen-phd}, the RPR-2007 model
\cite{corthals-2006,tamara-phd}, and Kaon-{\sc Maid}~\cite{mart-1999}
are also evaluated.

A conclusive statement with regard to the $M \approx 1900$~MeV
resonances is extremely useful to improve our understanding of the
nucleon's structure. Indeed, quark-diquark models do not predict a
resonance at this energy \cite{lichtenberg-1982a,ferretti-2011}. By
contrast, a number of resonances with a mass of around 1900 MeV is
predicted by {CQM}s \cite{capstick-2000,loring-2001-nonstrange}. 


\subsection{Likelihood function}
\label{subsec:likelihoodfunc}

When calculating the likelihood function against a single data set,
one usually does not take systematic errors $\sigma_\text{sys}$ into
account. However, this course of action is not valid when multiple
data sets are combined, as it would result in an underestimate of the
likelihood. Assuming that systematic errors are independent and
normally distributed, the total errors can be determined by adding the
systematic and statistical contributions in quadrature,
\begin{align}
 \sigma_\text{tot}^2 = \sigma_\text{stat}^2 + \sigma_\text{sys}^2 \; . 
\label{eqn:totalerror}
\end{align}
A more conservative estimate is to add the systematic and statistical
errors linearly
\begin{align}
 \sigma_\text{tot} ^{\prime} = \sigma_\text{stat} + \sigma_\text{sys} \; . 
\label{eqn:totalerrorconservative}
\end{align}
The numerical calculations for the Bayesian evidences are very
demanding and it is prohibitive to run the calculations with various
choices for the values of $\sigma _\text{tot}$.  In what follows, we
outline an approximate method which allows one to relate the evidences
computed with Eq.~(\ref{eqn:totalerror}) to those which use
Eq.~(\ref{eqn:totalerrorconservative}).

Most often, the systematic errors $\sigma_\text{sys}$ are computed by
taking the squared sum of a number of partial systematic errors
$\sigma_\text{sys}^{i}$ from different sources. This approach is prone
to underestimate the $ \sigma_\text{sys} $. For a systematic error
that is dominated by two errors with a comparable magnitude
$\sigma_\text{sys,1}\approx \sigma_{\text{sys},2}$, one obtains in the
conservative approach
\begin{align}
\sigma^{\prime}_\text{sys} = \sum_i \sigma_{\text{sys},i} 
\approx 2 \sigma_{\text{sys},1} \approx \sqrt{2}\sigma_\text{sys}.
\end{align}

In a scenario where $\sigma_\text{stat} \approx
\sigma_\text{sys}$, the estimate
(\ref{eqn:totalerror}) leads to $\sigma_{\text{tot}} \approx
\sqrt{2}\sigma_\text{stat}$ and to the following
value for a more conservative estimate of
$\sigma^{\prime}_{\text{tot}}$
\begin{align}
\sigma^{\prime}_\text{tot} = 
\sigma_\text{stat} + \sigma^{\prime}_\text{sys} 
\approx  \sigma_\text{stat} + \sqrt{2}\sigma_\text{sys}
\approx \frac{1+\sqrt{2}}{\sqrt{2}}\sigma_{\text{tot}}. 
\label{eqn:sigmatotprime}
\end{align}

\begin{figure}[tpb]
 \centering
\includegraphics[width=0.4\columnwidth]{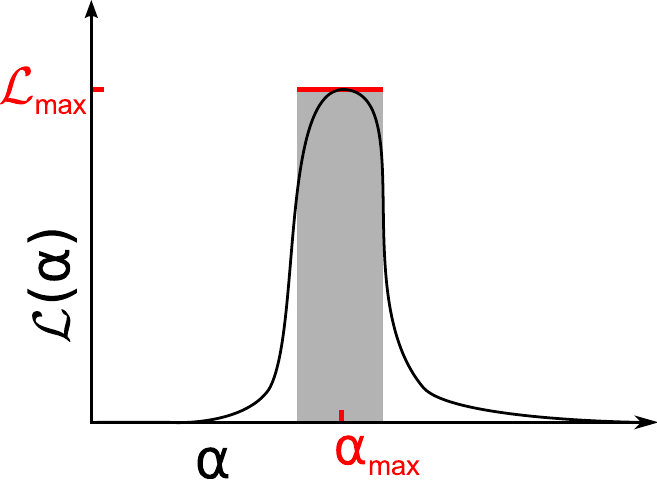}
 \caption{The integral over the likelihood function (black curve) can be approximated by the contribution for which $\m{L}(\alpha) \approx \m{L}_{\text{max}}$, or conversely, for which $\chi^2(\alpha) =  \chi^2_{\text{min}}$ (grey box).}
 \label{fig:scaleevidence}
\end{figure}

One can convert the $\m{Z}$ values computed with the errors of
Eq.~(\ref{eqn:totalerror}) into a $\m{Z}^{\prime}$ which use 
$\sigma^{\prime}_{\text{tot}}$. If the errors are multiplied by $c$,
the log-chi-square distribution $\ln \m{L}(\v{\alpha}_M)$ of
Eq.~(\ref{eqn:logchisquared}) scales as
\begin{align}
 S(k,\chi_R^2(\v{\alpha}_M),c)  &\equiv \ln \left(\dfrac{\m{L}(\v{\alpha}_M)}
{\m{L}^{\prime}_c (\v{\alpha}_M)}\right) \nonumber \\
 &= (k-2)\ln{c} - \chi^2_R \frac{k}{2}\;\frac{c^2-1}{c^2}.
\label{eqn:likelihood-scaling}
\end{align}
Here, $k$ denotes the number of degrees of freedom, $\chi_R^2
\equiv \chi^2/k$ is the reduced chi-squared as computed with the
values $\sigma_{\text{tot}}$. 
One can estimate the evidence resulting from the scaled likelihood function $\m{L}^{\prime}_c(\v{\alpha}_M)$ as follows.
Inserting a uniform prior into the Eq.~(\ref{eqn:full-evidence})
yields the following expression for $\m{Z}$
\begin{align}
\m{Z} = \dfrac{1}{\Delta} \int_{\v{\alpha}_{\text{0}}}^{\v{\alpha}_{\text{1}}} \m{L}(\v{\alpha})d\v{\alpha} \approx \dfrac{1}{\Delta} \int_{D} \m{L}(\v{\alpha}_{\text{max}})d\v{\alpha},
\end{align}
where $\Delta \equiv \prod_i \left( \Delta{\alpha}_i \right)$ is the
volume of the prior hypercube. Indeed, if $\m{L}(\v{\alpha})$ is the
chi-square distribution with $\chi_R^2(\v{\alpha})$ far from its
optimal value of 1 (\emph{e.g.} $\chi_R^2 = 4$), it falls rapidly with
increasing $\chi^2(\v{\alpha})$, and the bulk of the likelihood
originates from a volume $D$ where $\m{L}(\v{\alpha}) \approx
\m{L}_{\text{max}}$, or $\chi_R^2(\v{\alpha}) \approx
\chi^2_{R,\text{min}}$, as illustrated in
Fig.~\ref{fig:scaleevidence}.

By expressing the corrected likelihood $\m{L}^{\prime}_c(\v{\alpha})$
in terms of the likelihood $\m{L}(\v{\alpha})$ and the
scaling factor of Eq.~(\ref{eqn:likelihood-scaling}), the
expression for $\m{Z}^{\prime}$ becomes
\begin{align}
\m{Z^{\prime}} 	&\approx \dfrac{1}{\Delta} \int_{D} \m{L}^{\prime}_c(\v{\alpha_\text{max}})d\v{\alpha}\nonumber \\
		&\approx \dfrac{1}{\Delta} \int_{D} \m{L}(\v{\alpha}_\text{max}) e^{-S\left(k,\chi_R^2({\v{\alpha}_{\text{max}})},c \right)}\;d\v{\alpha}\\
		&\approx \m{Z}\;e^{-S\left(k,\chi^2_{R,\text{min}},c\right)},
\intertext{which yields our final result}
		\ln\m{Z}^{\prime} &\approx \ln \m{Z}-S\left(k,\chi^2_{R,\text{min}},c\right). 
\label{eqn:Zcorrection}
\end{align}
The expression~(\ref{eqn:totalerror}) for $\sigma_\text{tot}$
presupposes stringent independences between the various
contributions. The $\sigma_\text{tot} ^{\prime}$ of
Eq.~(\ref{eqn:totalerrorconservative}) provides a more conservative
estimate of the evidence. In order to avoid overestimating the amount
of information that is provided by the data, we will use
$\sigma_\text{tot} ^{\prime}$ in the forthcoming analyses.

\subsection{Identifying the resonance content of
  $p(\gamma,K^+)\Lambda$}
\label{subsec:resonancecontent}

\begin{figure}[tpbh]
 \centering
\includegraphics[width=.95\columnwidth]{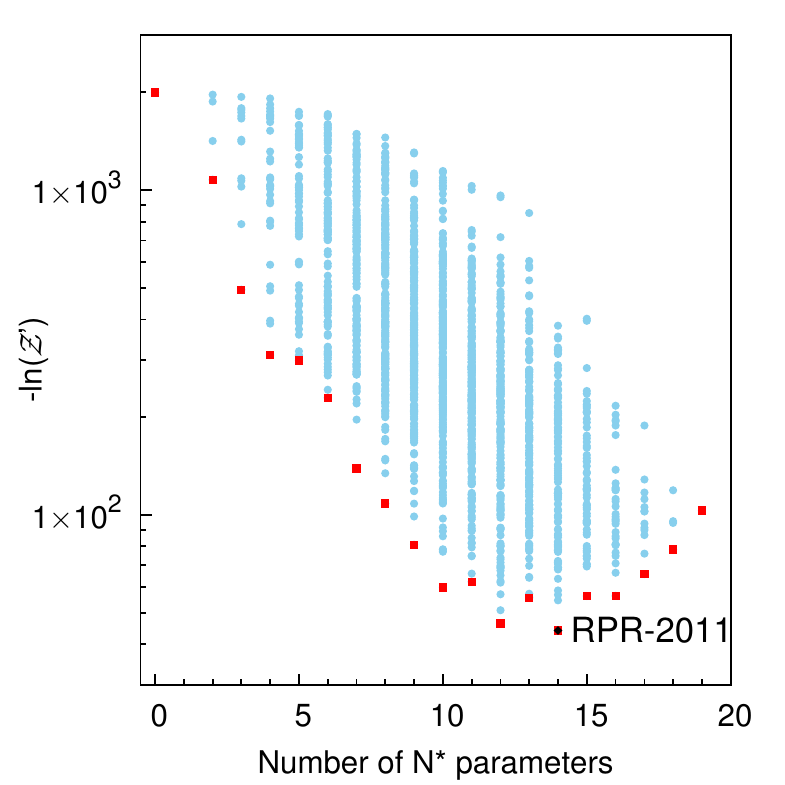}
\caption{(color online). The evidences $(-\ln
  {\mathcal{Z}^{\prime}})$ of the 2048 model variants in the RPR model
  space (blue circles), as a function of the number of free $N^{\ast}$
  parameters. The latter value equals the number of $N^{\ast}$
  couplings plus one for the $\Lambda _ {R}$. The smaller the value of
  $-\ln\m{Z}^{\prime}$ the higher the evidence. The best model for a
  fixed number of parameters is indicated with a red square. The model
  with the highest evidence, RPR-2011, is denoted with a black
  diamond.}
 \label{fig:bestrpr2011}
\end{figure}

The 11 proposed resonances of Table~\ref{tab:resonances} give rise to
$2^{11} = 2048$ model variants. The Bayesian evidence $\m{Z}^{\prime}$
of Eq.~(\ref{eqn:Zcorrection}) is computed for each model, resulting
in a map of the RPR model space, shown in Fig.~\ref{fig:bestrpr2011}.

\begin{figure*}[tbh]
	\includegraphics[width=.65\textwidth]{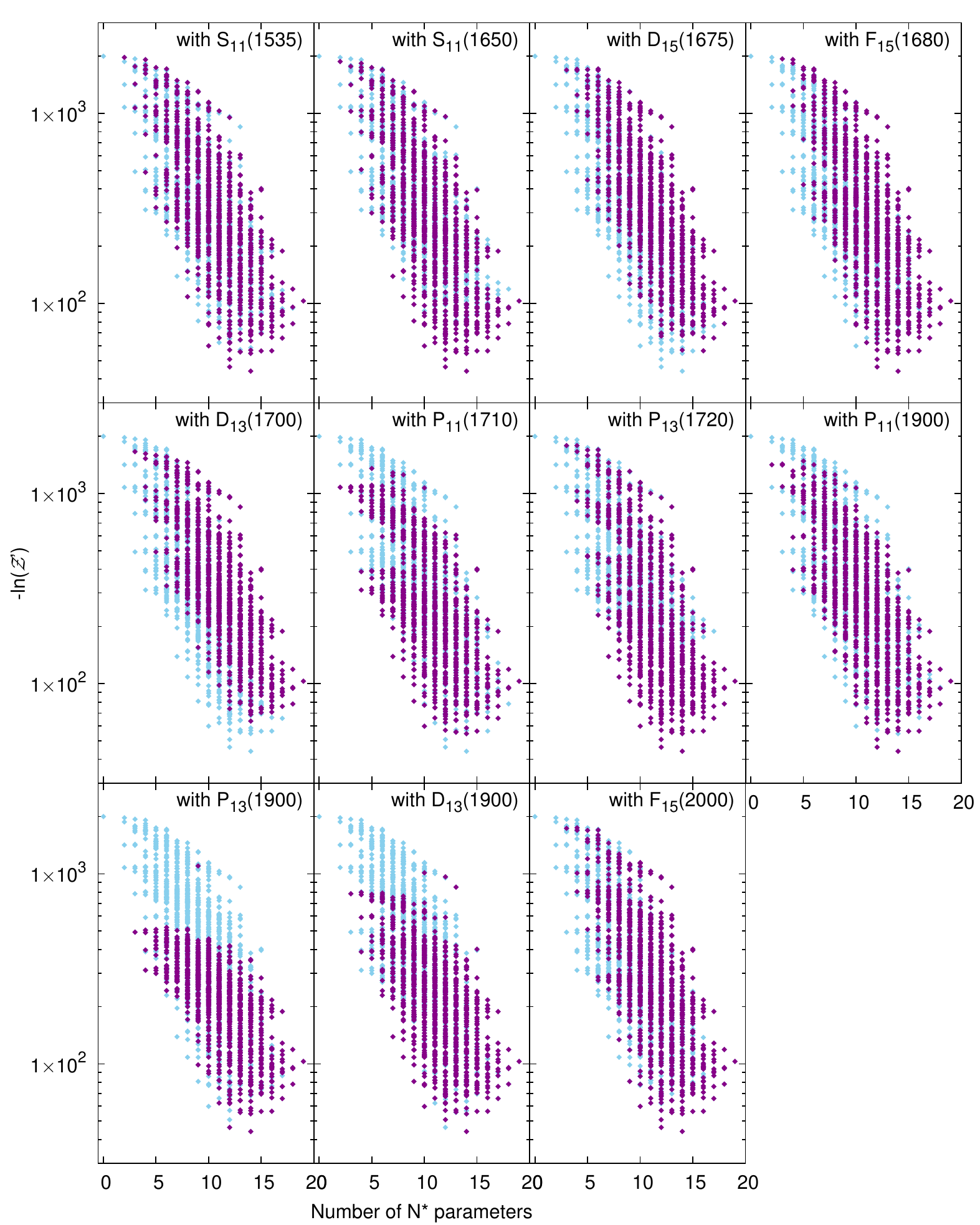}
	\caption{(color online). The evidences $(-\ln
          {\mathcal{Z}^{\prime}})$ of the 2048 model variants in the
          RPR model space (blue circles). The purple diamonds
          correspond with the subset of models which contain the
          resonance indicated in the top right corner of each panel.}
	\label{fig:lnz-res-all}
\end{figure*}
\begin{figure*}[tbh]
	\includegraphics[width=.65\textwidth]{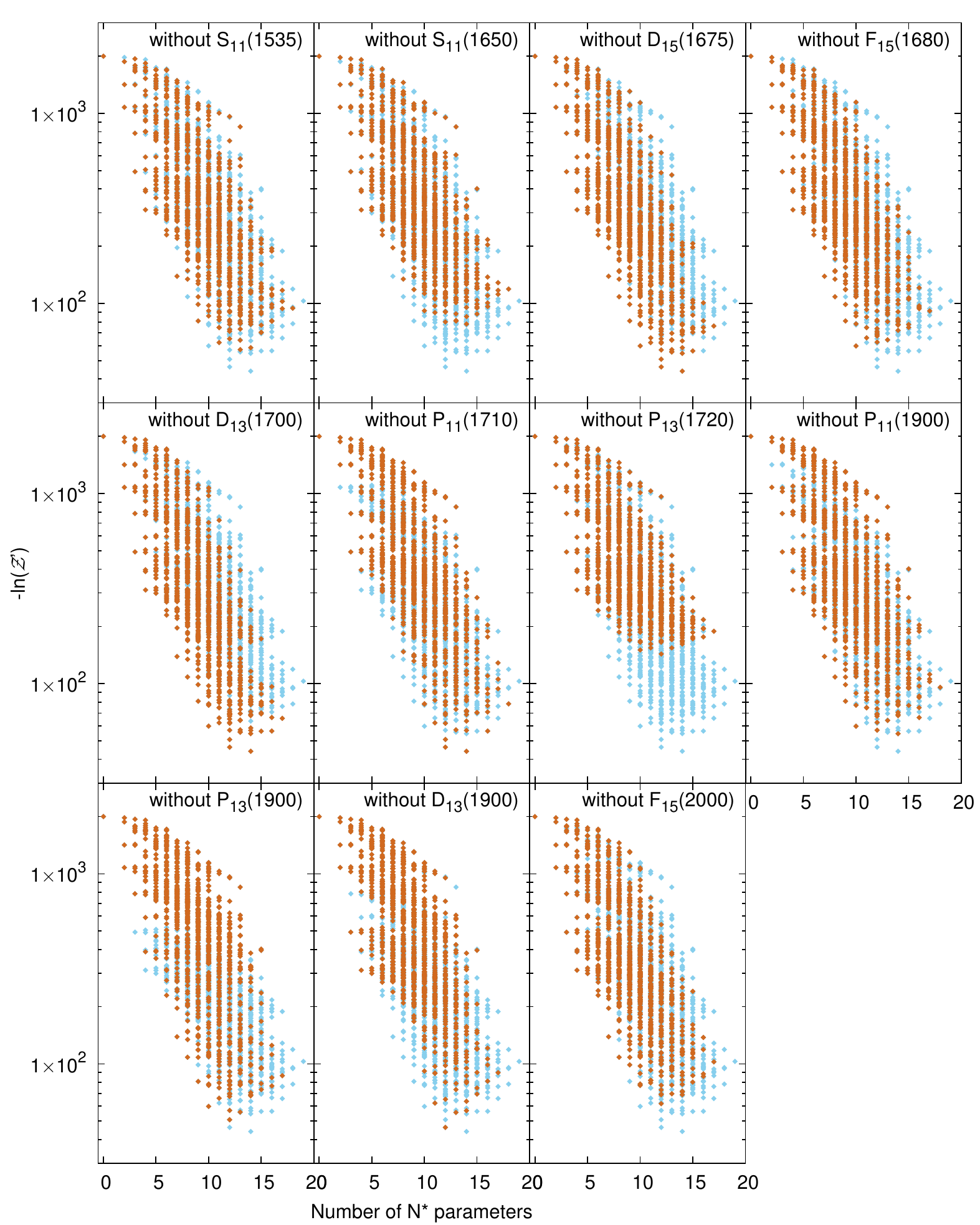}
	\caption{(color online). The evidences $(-\ln
          {\mathcal{Z}^{\prime}})$ of the 2048 model variants in the
          RPR model space (blue circles). The orange diamonds
          correspond with the subset of models which do not contain
          the resonance indicated in the top right corner of each
          panel.}
	\label{fig:lnz-nores-all}
\end{figure*}

The parameters of the Reggeized background are assigned localized
priors of 20\% around the values determined in
Section~\ref{sec:bg}. Therefore, the total number of adjustable
parameters is the sum of the number of $N^{\ast}$ couplings, of the
three background parameters, and of the cut-off value $\Lambda_ {R}$
of Eq.~(\ref{eq:valueofcutoff}).  The number of fitted $N^{\ast}$
couplings extends from 1 (one spin-1/2 coupling) to 18 ($4\times 1$
spin-1/2, $4 \times 2$ spin-3/2, $3 \times 2$ spin-5/2 couplings). We
adopt one common value for $\Lambda_ {R}$ for all resonances with a
uniform prior between 1.0 and 3.5~GeV.  When selecting a prior
distribution, it is good practice to ignore the data. Often an
overestimation of the evidence results from determining the likelihood
and the prior with a particular data set. In this work, the ranges of
the prior distributions of the resonance couplings are selected on the
basis of naturalness arguments.  Indeed, the contribution of a single
resonance is unlikely to exceed the total $p(\gamma,K^{+})\Lambda$
cross section ($\sigma \approx 5\mu b$) by a large factor. We
performed calculations of the total cross sections ($\sigma_R$) in a
model which includes a single resonance $R$ and the Reggeized
background. It is observed that the criterion $\sigma_R < 25\mu b$
leads to absolute values of the coupling constants smaller than 100 in
the adopted units convention. Therefore, we adopt a uniform
distribution for the priors of the resonance coupling constants in the
range $[-100,100]$.

Jeffreys' scale allows us to determine the ``best'' model from the
2048 variants. The model with the highest evidence has $14$ $N^{\ast}$
parameters (13 couplings and $\Lambda_{R}$) and features the
$S_{11}(1535)$, $S_{11}(1650)$, $F_{15}(1680)$, $P_{13}(1720)$,
$P_{11}(1900)$, $F_{15}(2000)$, and the missing $D_{13}(1900)$ and
$P_{13}(1900)$. This model variant will be referred to as RPR-2011 
\cite{DeCruz:2011xi}.
The ``second-best'' model has two parameters less due to the absence
of the $D_{13}(1900)$.  The difference in $-\ln\m{Z}$ between the
``best'' and ``second-best'' models is 2.3. This corresponds to
significant to strong evidence in favor of RPR-2011. The difference
with the other models is at least 6.8, which is consistent with
decisive evidence for RPR-2011.

\begin{figure}[bh]
	\includegraphics[width=\columnwidth]{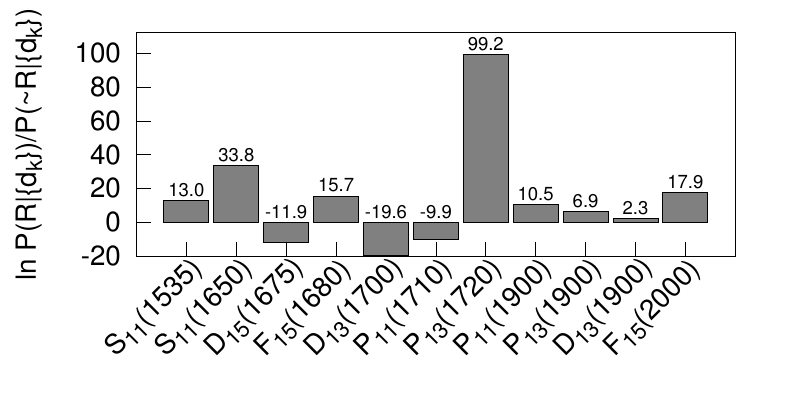}
	\caption{The relative resonance probabilities of
          Eq.~\ref{eq:propabilityratios} for each resonance listed in
          Table~\ref{tab:resonances}.}
	\label{fig:estimateratio}
\end{figure}

In a next step, one can quantify the probability of each resonance
separately by evaluating $P\left(R \; | \left\lbrace d_k \right\rbrace
\right)$ of Eq.~(\ref{eqn:bayesapp1}). Fig.~\ref{fig:lnz-res-all}
visualizes which models are included in the sum. It is also
instructive to calculate $P\left(\sim R \; | \left\lbrace d_k
  \right\rbrace \right)$, the probability that a resonance is
\emph{not} required to describe the reaction. The calculation of this
quantity is completely analogous to Eq. (\ref{eqn:bayesapp1}). In
Fig.~\ref{fig:lnz-nores-all}, the models which do not include a
resonance $R$ are visualized for each proposed resonance.

The probability ratios 
\begin{equation}
\ln \left( P\left(R \; | \left\lbrace
    d_k \right\rbrace \right)/P\left(\sim R \; | \left\lbrace d_k
  \right\rbrace \right) \right), 
\label{eq:propabilityratios}
\end{equation}
are plotted in Fig.~\ref{fig:estimateratio}.
A positive ratio indicates that the probability that the resonance $R$
contributes to the reaction $p(\gamma,K^+)\Lambda$ is greater than the
probability that it does not. Conversely, a negative ratio means that
the data does not support the possibility that $R$ contributes to the
reaction.

The results indicate that the resonances in RPR-2011 are those that
have a positive probability ratio. Moreover, the two resonances with
the highest probabilities are $P_{13}(1720)$ and $S_{11}(1650)$. These
are the two resonances that are also deemed important for the
description of $p(\gamma,K^+)\Lambda$ by most other models.  There is
decisive evidence that the $D_{15}(1675)$, $D_{13}(1700)$, and
$P_{11}(1710)$ are not required to describe the $p(\gamma,K^+)\Lambda$
data.  With regard to the resonance content in the 1800-2000~MeV mass
range several suggestions have been in the literature, but no
consensus has been reached.  We have evaluated four states in that
mass region (Table~\ref{tab:resonances}): two have a two-star status
and two are labeled as ``missing''. Our analysis provides decisive
evidence for three of these states: $P_{13}(1900)$, $P_{11}(1900)$,
and $F_{15}(2000)$. Note that the evidence for the ``missing''
$D_{13}(1900)$ is significant to strong, but not decisive.

\begin{figure*}[tbh]
	\includegraphics[scale=0.65]{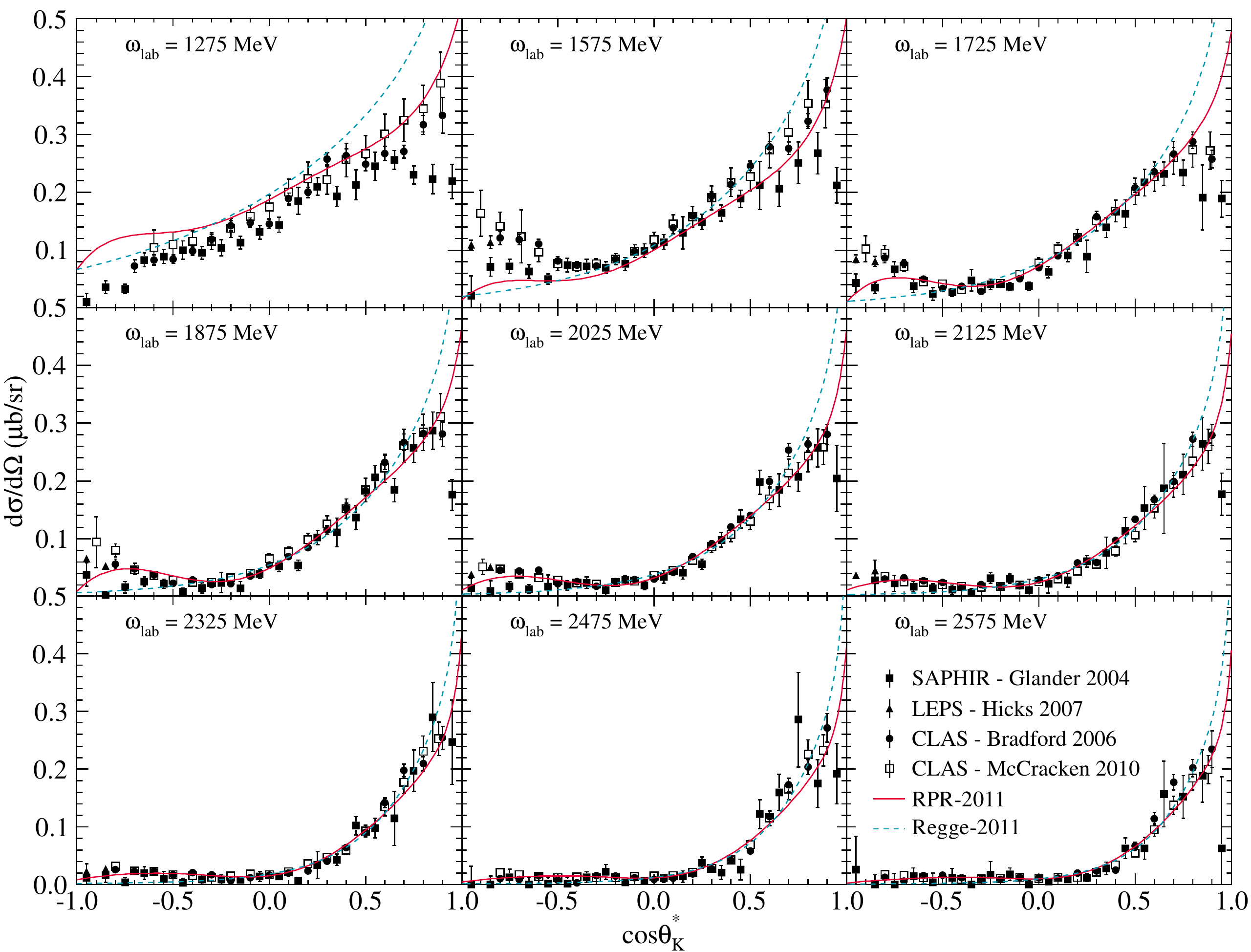}
	\caption{(color online). Angular dependence of the
          differential cross section at various incident photon
          energies $\omega_{\text{lab}}$. The full red line represents
          the RPR-2011 model, the blue dashed line corresponds with
          Regge-2011. Data are from
          Refs. \cite{glander-2004,hicks-2007,bradford-2006,mccracken-2009}. }
	\label{fig:dcs}
\end{figure*}

\begin{figure*}[tbh]
	\includegraphics[scale=0.65]{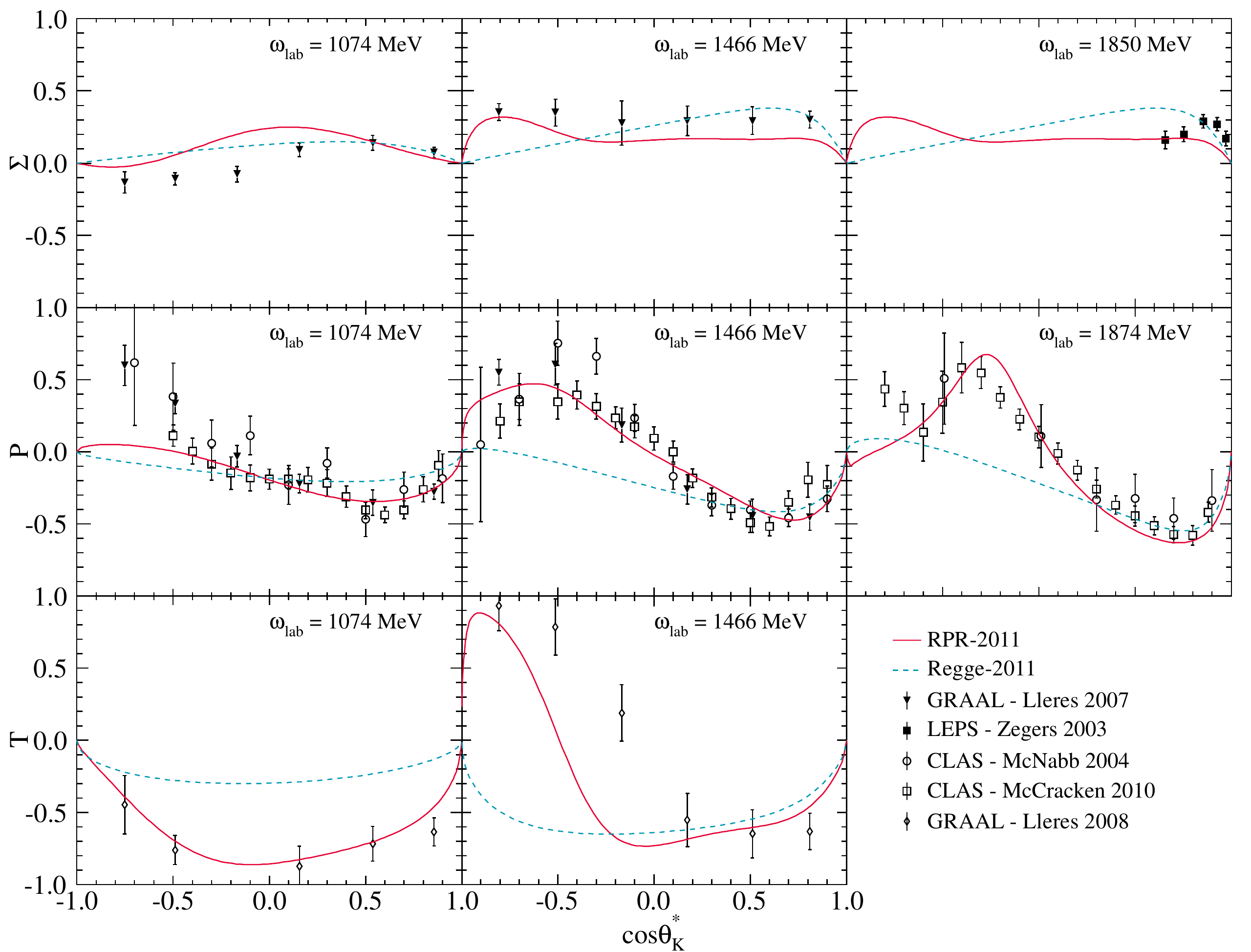}
	\caption{(color online). Angular dependence of the single polarization observables: beam asymmetry $\Sigma$ (top row), recoil polarization $P$ (middle row), and target asymmetry $T$ (bottom row), at various incident photon energies $\omega_{\text{lab}}$. Line conventions as in Fig.~\ref{fig:dcs}. Data are from Refs. \cite{zegers-2003,mccracken-2009,mcnabb-2004,lleres-2008,lleres-2007}. }
	\label{fig:sgpol}
\end{figure*}
\subsection{Photoproduction with RPR-2011}
\label{subsec:photonresults}

The $p(\gamma,K^+)\Lambda$ observables presented in this subsection
are calculated with the RPR-2011 model parameters fixed at their
maximum likelihood values. The RPR-2011 results are compared to the
Reggeized background model Regge-2011 that was determined in
Section~\ref{sec:bg}. From the difference between the Regge-2011 and
the RPR-2011 results one can infer conclusions about the role of the
resonances for the various observables.

The $p(\gamma,K^{+})\Lambda$ differential cross section is displayed
as a function of $\cos\theta_K^{\ast}$ in Fig.~\ref{fig:dcs}. RPR-2011
provides a good description over a wide range of kinematics.  For the
lowest energies and backward angles there are deviations between the
model and the data, hinting at possible missing dynamics such as
$u$-channel contributions.  It is striking that the $t$-channel
background of the Regge-2011 model already provides a reasonable
description of the gross features of both the $\omega_{lab}$ and $\cos
\theta_{K} ^{\ast}$ dependence of the differential cross sections. The
biggest effect from the resonance contributions is observed at the
forward and backward kaon angles.

Both RPR-2011 and Regge-2011 models exhibit a steep rise at extremely
forward angles. At the three lowest $\omega_{lab}$ energies considered
in Fig.~\ref{fig:dcs} the inclusion of the resonances softens this
rise and improve the goodness of the fit to the data.  Note that the
SAPHIR data (Fig.~\ref{fig:dcs__bin-wlab__func-cos_RPR2007}) suggest a
plateau at forward kaon angles and that this feature is absent in the
CLAS data. 


The angular dependence of the single-polarization observables
$\Sigma,P,T$ is shown for three representative energies in
Fig.~\ref{fig:sgpol}.  The $\Sigma,P,T$ receive stronger contributions
from the $N ^{\ast}$'s than the differential cross sections. In contrast
to the high-energy situation considered in Fig.~\ref{fig:pol-iso1-RR},
the photon asymmetries in the resonance region are relatively
small. The Regge-2011 reproduces the trend of increasing $\Sigma $
with growing $\omega_{lab}$. The inclusion of the $N^{\ast}$'s does not
lead to a considerably improved quality of the fit.  The recoil
polarization $P$ and target polarization $T$ are highly sensitive to
the resonance contributions. 

\begin{figure*}[tbh]
	\includegraphics[scale=0.65]{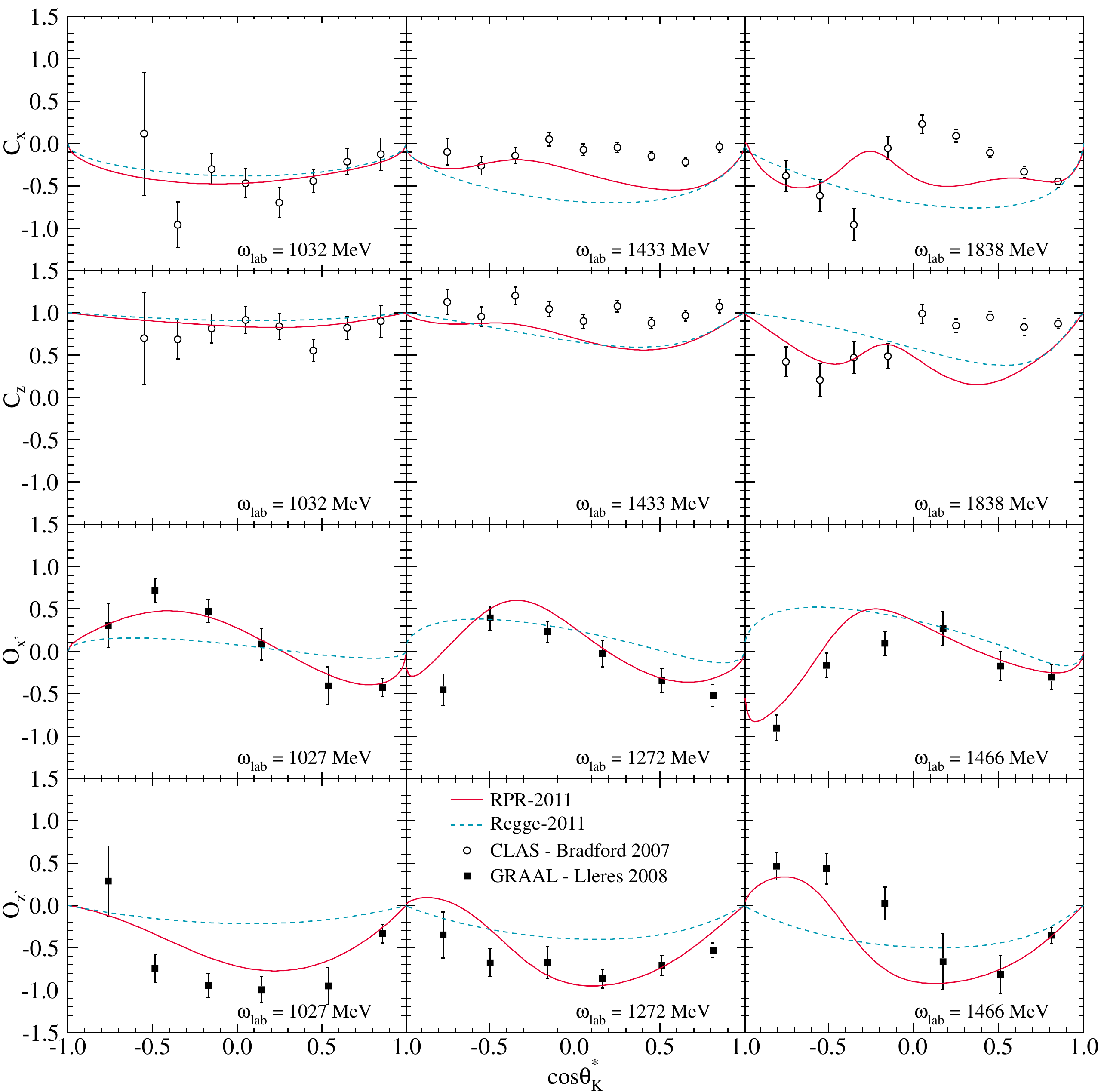}
	\caption{(color online). Angular dependence of the beam-recoil double polarization observables with circular beam polarization, $C_x$ and $C_z$ (top rows), and with oblique beam polarization, $O_{x^{\prime}}$ and $O_{x^{\prime}}$ (bottom rows), at various incident photon energies $\omega_{\text{lab}}$. Line conventions as in Fig.~\ref{fig:dcs}. Data are from Refs. \cite{lleres-2008,bradford-2007}.}
	\label{fig:dbpol}
\end{figure*}

We now turn our attention to the double polarization observables. We
stress that they represent but 7\% of the total amount of data and
that we give each data point an equal weight.  As the bulk of the data
is in the differential cross sections and to a lesser extent in the
single polarization observables, the double polarization observables
represent stringent tests of the RPR-2011 model.  Perhaps somewhat
surprisingly, Regge-2011 provides a good approximation to the
double-polarization observables $C_x$ and $C_z$. The observed trends
$C_z \approx 1$ and $C_x \approx C_z - 1$ \cite{schumacher-2008} are
well reproduced by both the Reggeized background (Regge-2011) and
RPR-2011. This observation hints at the fact that the $C_x, C_z$ are
very background dominated.  A large sensitivity to resonance
contributions is observed for $O_{x^{\prime}}$ and $O_{x^{\prime}}$,
which are considerably better described by RPR-2011 than by
Regge-2011.

\begin{figure}[tbp]
\centering
\includegraphics[width=\columnwidth]
{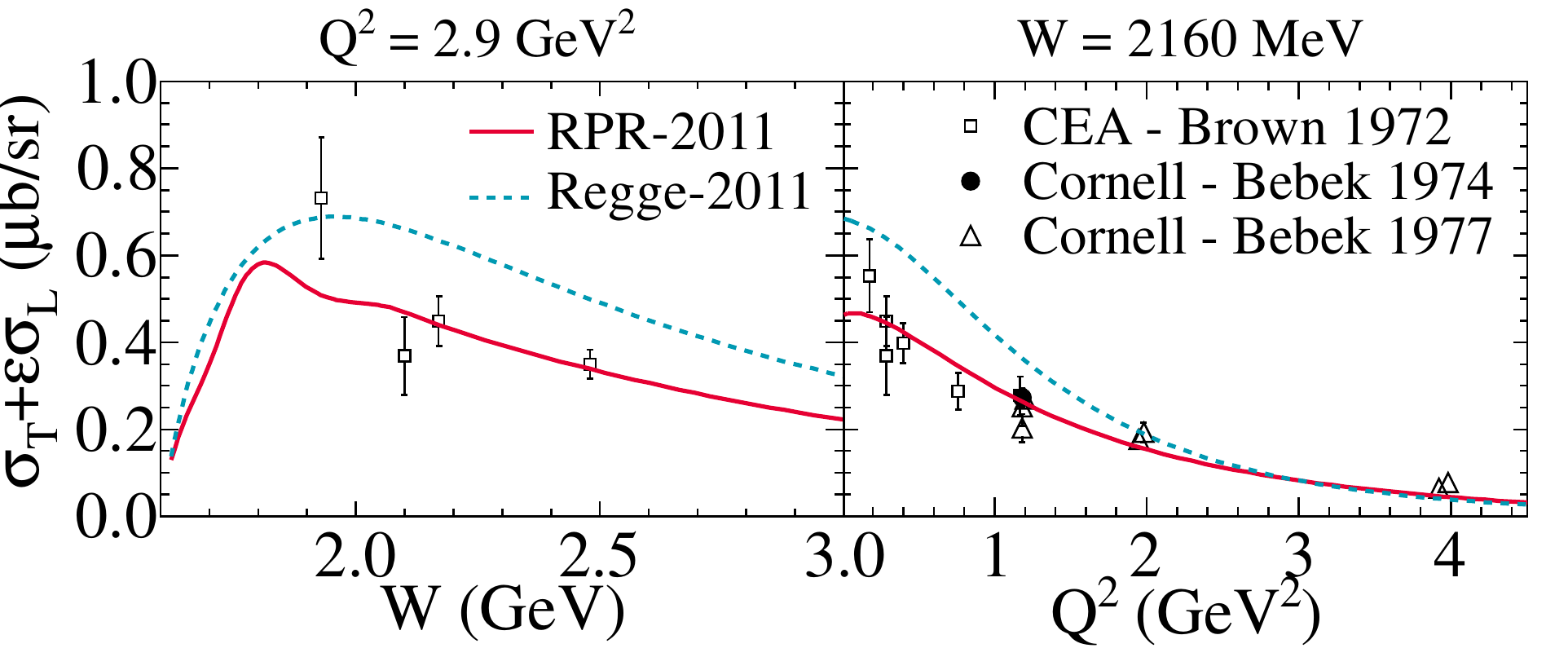}
 \caption{The unseparated structure functions $\sigma_T +
   \varepsilon\, \sigma_L$ for $p(e,e^{\prime}K^{+})\Lambda$ at
   $\cos\theta_K^{\ast} \approx 1$ as a function of $W$, at $Q^2 = 2.9
   $ GeV$^2$ (left panel) and as a function of $Q^2$, at $W = 2160$
   MeV (right panel).  Line conventions as in Fig.~\ref{fig:dcs}. Data
   are from Refs. \cite{brown-1972,bebek-1974,bebek-1977}.}
\label{fig:diff-l+t}
\end{figure}

\begin{figure}
 \centering
 \includegraphics[width=\columnwidth]{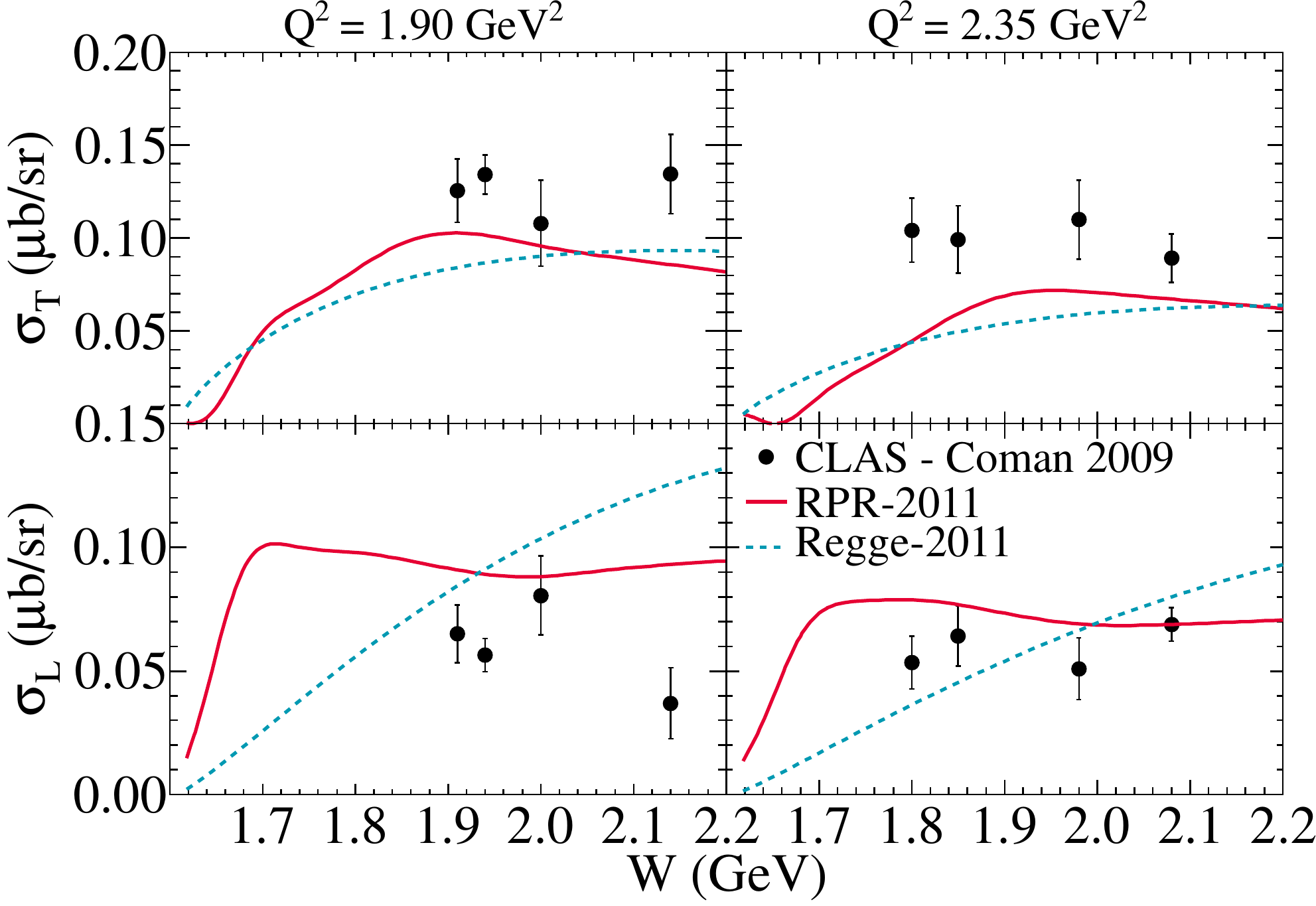}
 \caption{The separated structure functions $\sigma_T$ (top panels)
   and $\sigma_L$ (bottom panels) for $p(e,e^{\prime}K^{+})\Lambda$ at
   $\cos\theta_K^{\ast} \approx 1$ as a function of $W$, at $Q^2 =
   1.90 $ GeV$^2$ (left panels) and at $Q^2 = 2.35 $ GeV$^2$ (right
   panels).  Line conventions as in Fig.~\ref{fig:dcs}. Data are from
   Ref.~\cite{coman-2009}.}
 \label{fig:rpr2011-diff-l_t_coman}
\end{figure}

\subsection{Electroproduction predictions with RPR-2011}
\label{subsec:electronresults}

Electroproduction reactions have the potential to probe the hadron's
electromagnetic substructure. The existence of this substructure can
be parametrized by introducing a $Q^2$ dependence in the
electromagnetic coupling constants. In addition, the
$p(e,e^{\prime}K^{+})\Lambda$ reaction dynamics becomes sensitive to
longitudinal couplings.  In part, these couplings arise naturally from
the photoproduction amplitudes when gauge invariance is imposed.  A
peculiar class of longitudinal couplings vanishes for photoproduction
reactions and cannot be constrained against real-photon data. The
latter class of longitudinal couplings are neglected in the RPR
model. In brief, we fix the basic reaction mechanism to the
$p(\gamma,K^{+})\Lambda$ data and treat the electroproduction data as
a test of the model. Thereby, we make some reasonable assumptions with
regard to the electromagnetic form factors of the $t$-channel kaons
and $s$-channel $N^{\ast}$'s. Such an extrapolation of the
$p(\gamma,K^{+})\Lambda$ amplitude to $p(e,e^{\prime}K^{+})\Lambda$
has been shown as reasonably successful for the RPR-2007 model
\cite{corthals-2007b}. We use the same $N ^{\ast}$ helicity amplitudes
(HA) as in Ref.~\cite{corthals-2007b}. Also the transition form
factors for the spin-$1/2$ particles are those from
Ref.~\cite{corthals-2007b}. The transition form factors for the
spin-$3/2$ particles are derived from the consistent Lagrangians of
Eqs.~(\ref{eqn:LEM32-1})--(\ref{eqn:LEM32-2}). For the spin $3/2$ and
$5/2$ particles, the HA are calculated in the Bonn constituent quark
model~\cite{loring-2001-nonstrange}, and the transition form factors
are derived using Eqs.~(\ref{eqn:LEM52-1})--(\ref{eqn:LEM52-2}).

A comparison between recent low $Q ^2=0.030-0.055$~GeV$^2$
measurements and RPR-2011 predictions are contained in
Ref.~\cite{achenbach-2011}.  It was observed that RPR-2011 provides a
fair description of those data.  Unseparated structure functions
$\sigma_T + \varepsilon\, \sigma_L$ at very forward kaon angles
obtained in the 1970s are shown as a function of $W$ and $Q^2$ in
Fig.~\ref{fig:diff-l+t} together with Regge-2011 and RPR-2011
predictions.  Obviously, at $\cos\theta_K^{\ast} \approx 1$ the major
impact of the intermediate resonances is to reduce the cross section
by some modest factor. This is in line with the observations made for
the real-photon differential cross sections of Fig.~\ref{fig:dcs}. The
electromagnetic form factors of the intermediate resonances reduce the
effect of the $N ^{\ast}$'s with growing photon virtuality $Q ^{2}$.
The RPR-2011 model provides a fair prediction for both the $Q ^{2}$
and $W$ dependence of the data.

Fig.~\ref{fig:rpr2011-diff-l_t_coman} shows the energy dependence of
the separated structure functions $\sigma_L$ and $\sigma_T$.  In line
with the data, RPR-2011 predicts a $\sigma _{L}$ and $\sigma _{T}$ of
almost equal magnitude.  The $\sigma_T$ appears to be systematically
underpredicted while $\sigma_L$ is somewhat overpredicted. The fair
reproduction of both the magnitude and the $W$ dependence of
$\sigma_{L}$ provides support for our assumptions with regard to the
longitudinal couplings.

%

\begin{figure}[tbp]
 \centering
    \includegraphics[width=\columnwidth]{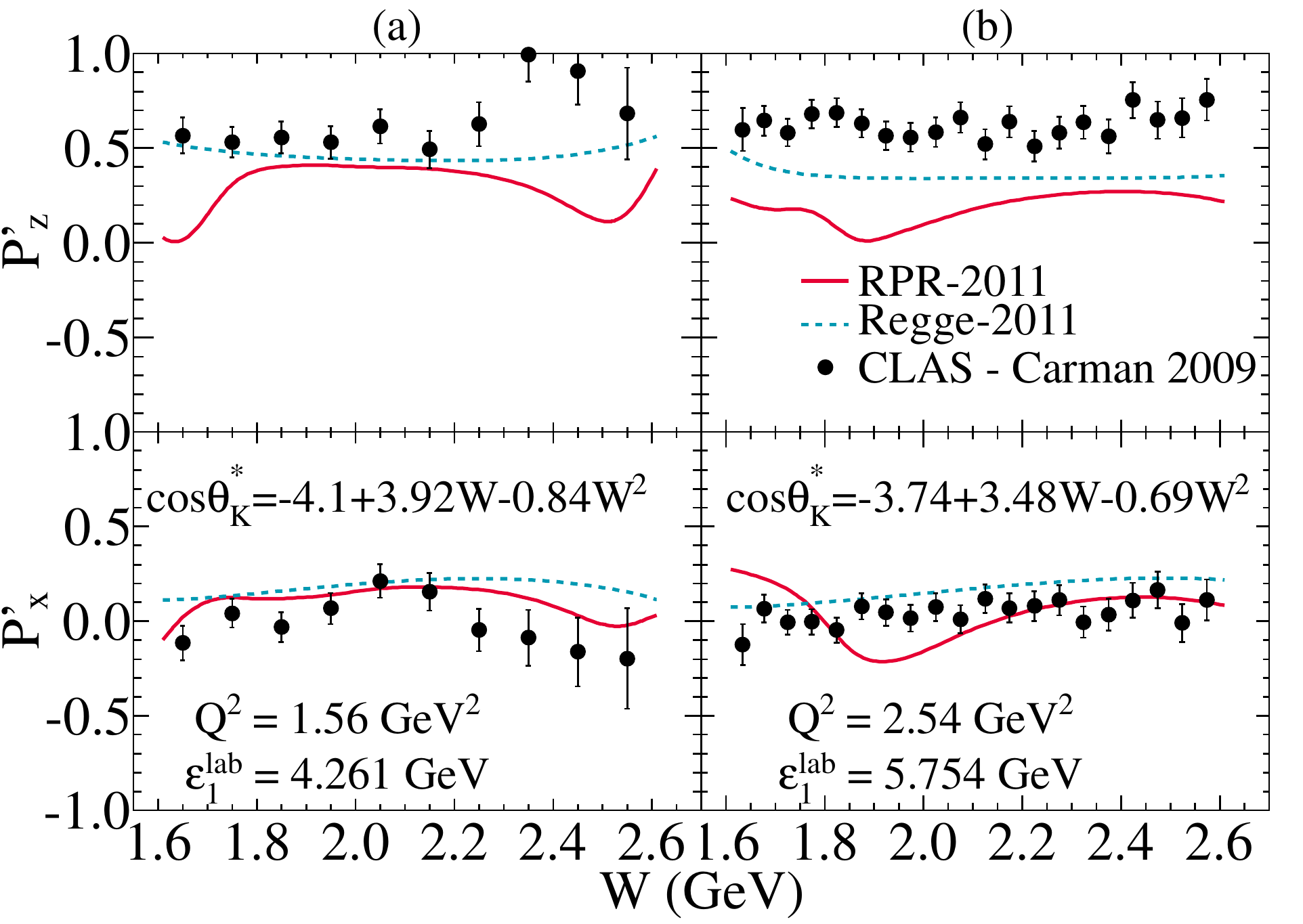}
    \caption{Transferred polarisation $P^{\prime}_z$ (top panels) and
      $P^{\prime}_x$ (bottom panels) as defined in Table~III of Ref.~\cite{carman-2009} for
      $p(\vec{e},e^{\prime}K^{+})\vec{\Lambda}$ at (a) $Q^2=1.56$ GeV$^2$
      and $\epsilon _{1} ^{lab}$=4.261~GeV (a), and (b) $Q^2=2.54$
      GeV$^2$ and $\epsilon _{1} ^{lab}$=5.754 GeV.  The $W$ dependence of  
      $\cos\theta_K^{\ast}$ can be captured by the following functions:
      (a) $\cos\theta_K^{\ast} = -4.1 + 3.92 W - 0.84 W^2$, and
      (b) $\cos\theta_K^{\ast} = -3.74 + 3.48 W - 0.69 W^2$.  Line
      conventions as in Fig.~\ref{fig:dcs}.  Data are from
      Ref.~\cite{carman-2009}.}
 \label{fig:rpr2011-transfpol}
\end{figure}

Predictions for the transferred polarisation are presented in
Fig.~\ref{fig:rpr2011-transfpol}.  The Reggeized background model
Regge-2011 as it was determined in
Sect.~\ref{subsec:optimumbackground} predicts the flat $W$ dependence
and the magnitude of $P^{\prime}_z \approx 0.0 $ and $P^{\prime}_x
\approx 0.5 $. For $P^{\prime}_z$, the introduction of resonances
worsens the quality of the agreement with the data obtained in
Regge-2011.  For $P^{\prime}_x$ the effect of the $N ^{\ast}$ is
smaller and the quality of the agreement is better than for
$P^{\prime}_z$.

\section{Conclusion} 
\label{sec:conclusion} 

The RPR framework conjoins a Reggeized $t$-channel background with
tree-level $s$-channel nucleon resonances from an isobar approach
into an economical model for kaon photoproduction in and above the
resonance region. The RPR model clearly separates nonresonant from
resonant amplitudes which is an asset when searching for the
properties of those (missing) resonances which contribute to
$p(\gamma,K^+)\Lambda$. 

We have used Bayesian inference to perform model selection both with
regard to the resonant and nonresonant content of the RPR
framework. It was shown that the Bayesian evidence $\mathcal{Z}$ is a
quantitative measure for a model's fitness given data. The
computation of $\mathcal{Z}$ requires involving multidimensional
integrals which demand dedicated numerical methods. To that
purpose we have proposed the ``GA+{\sc minuit}+log-{\sc vegas}''
integration strategy. With this method one can reliably compute 
$\mathcal{Z}$ for models with a moderate number of adjustable
parameters such as the RPR framework.

First, the most probable model variant for the Reggeized background
was determined against the 2.6~GeV$<W<$3.0~GeV data. This involves the
determination of three continuous and two discrete adjustable
parameters. The extracted value for $g_{K ^{+} \Lambda p}$ is
compatible with the one predicted by SU(3) symmetry.  Next, we have
considered a set of 11 nucleon resonances to determine the optimum
resonant contribution in the RPR $p(\gamma,K^+)\Lambda$ framework. To
this end, the Bayesian evidence was calculated for all 2048 model
variants resulting from the various resonance combinations. The model
with the highest evidence, dubbed RPR-2011, includes the resonances
$S_{11}(1535)$, $S_{11}(1650)$, $F_{15}(1680)$, $P_{13}(1720)$,
$P_{11}(1900)$, $F_{15}(2000)$, $D_{13}(1900)$, and $P_{13}(1900)$. An
evaluation of the individual resonances' probabilities reveals that
the two resonances with the highest evidence of contributing to
$p(\gamma,K^+)\Lambda$ are the $S_{11}(1650)$ and
$P_{13}(1720)$. There is decisive evidence that the $D_{15}(1675)$,
$D_{13}(1700)$, and $P_{11}(1710)$ are not required to describe the
current $p(\gamma,K^+)\Lambda$ world's data. The computed evidence for
the two-star $P_{13}(1900)$, the two-star $F_{15}(2000)$, and the
``missing'' $P_{11}(1900)$ is decisive, whereas for the ``missing''
$D_{13}(1900)$ it is significant, but not decisive.

After fixing the basic reaction reaction mechanism to the
$p(\gamma,K^{+})\Lambda$ data, the electroproduction data serve as
a test of the model.  In general our predictions for the
electroproduction data are reasonably good which proves that the
RPR-2011 model possesses predictive power and goes beyond a mere
analysis framework. Therefore, we consider RPR-2011 as an efficient
and robust model which can, for example, be used as an elementary production
operator in strangeness production reactions involving the deuteron 
and finite nuclei.

\begin{acknowledgments}
This research was financed by the Flemish Research Foundation (FWO Vlaanderen). The computational resources (Stevin Supercomputer Infrastructure) and services used in this work were provided by Ghent University, the Hercules Foundation and the Flemish Government -- department EWI.
\end{acknowledgments}


\bibliography{kl2012}

\end{document}